\documentclass[a4paper,12pt]{article}

\pdfoutput=1 

\usepackage[hmargin=.71in,vmargin=1.1in]{geometry}
\usepackage{indentfirst}
\usepackage{amsfonts}
\usepackage{mathrsfs}
\usepackage{amsthm}
\usepackage{amsmath}
\usepackage{mathabx}
\usepackage{amssymb}
\usepackage{authblk}
\usepackage{cite}
\usepackage{xcolor}
\usepackage{mathtools}
\usepackage{tensor}
\usepackage{graphicx}
\usepackage{bm}
\usepackage{upgreek}
\usepackage{braket}
\usepackage{color,soul}
\usepackage{csquotes}
\usepackage{caption}
\usepackage{subcaption}
\usepackage{slashed}
\usepackage{mathtools}
\usepackage{soul}
\usepackage{bbold}
\usepackage{booktabs}

\usepackage[bookmarksnumbered=true,bookmarksopen=true]{hyperref}
 \hypersetup{colorlinks,
             linkcolor=[rgb]{0,0.3,0.6}, 
             citecolor=[rgb]{0,0.3,0.6}, 
             urlcolor=[rgb]{0,0.3,0.6}}

\theoremstyle{plain}

\theoremstyle{definition}

\theoremstyle{remark}

\usepackage{tikz}
\usetikzlibrary{matrix}

\begin{document}

\title{\Large\textbf{Environment-induced topological reconstruction of black-hole thermodynamics}}

\author[a]{Xinyu Zhou\thanks{zhouxinyu0735@s.ytu.edu.cn}}
\author[a]{Chen Lan\thanks{stlanchen@126.com}}
\author[b,c,d,e]{Zhen-Ming Xu\thanks{zmxu@nwu.edu.cn}}

\affil[a]{\normalsize{\em Department of Physics, Yantai University, 30 Qingquan Road, Yantai 264005, China}}
\affil[b]{\normalsize{\em School of Physics, Northwest University, Xi'an 710127, China}}
\affil[c]{\normalsize{\em Peng Huanwu Center for Fundamental Theory, Xi'an 710127, China}}
\affil[d]{\normalsize{\em Shaanxi Key Laboratory for Theoretical Physics Frontiers, Xi'an 710127, China}}
\affil[e]{\normalsize{\em Fundamental Discipline Research Center for Quantum Science and technology of Shaanxi Province, Xi'an 710127, China}}

\date{ }

\maketitle

\begin{abstract}
Thermodynamic topology describes black-hole (BH) equilibria as zeros, or defects, of a vector field built from the off-shell free energy.
Studies with surrounding matter find that the topological class is unchanged, but not how the matter's spatial distribution affects individual equilibria.
We ask whether the position of a localized environment can reorganize these equilibria at fixed total topological number.
We model the environment as a radially localized energy bump peaked at $a_m$ and add its exterior energy to the generalized off-shell free energy.
As $a_m$ varies, defects migrate and are created or annihilated in opposite-charge pairs, a process we call topological reconstruction.
It changes the number and stability of equilibrium branches but preserves the total topological number, provided no zero crosses the boundary of the physical domain.
We find this in Schwarzschild, Schwarzschild-AdS, Reissner-Nordstr\"om, and Reissner-Nordstr\"om-AdS BHs, with critical positions and required strength that depend on the background.
In our examples, the required environmental energy ranges from below $1\%$ of the BH mass to a value comparable to it.
The model is an effective fixed-background description without backreaction.
Within it, the total topological number cannot capture the environmental response; local defects must be tracked.
\end{abstract}

\tableofcontents

\section{Introduction}
\label{sec:introduction}
 
Black-hole (BH) thermodynamics links the equilibrium properties of gravitational systems to temperature, entropy, and free energy, and it underlies our understanding of BH stability and phase transitions~\cite{Hawking:1976de,Hawking:1982dh,Kubiznak:2016qmn,Witten:2024upt,Mann:2025xrb,Mann:2026nqs}.
Real BHs are not isolated.
Their thermodynamics has been examined in the presence of matter shells, cavities, and dark-matter halos~\cite{Brady:1991np,Katz:1994gz,Parentani:1994wr,Andre:2019zzo,Myung:2025cxu}, and the effects of accreting matter and dark matter on ringdown and other perturbative responses have been widely studied~\cite{Leung:1997was,Cardoso:2019rvt,Barausse:2014tra,Zhang:2022roh,Zhao:2023tyo,Liu:2024xcd,Daghigh:2023ixh,Chen:2023akf,Courty:2023rxk,Toshmatov:2025rln,Yuan:2025fde}.
In quasinormal-mode and echo studies, for example, the environment is often modeled as a localized bump in the exterior potential~\cite{Cardoso:2019rvt,Courty:2023rxk}.
These results do not directly determine how thermodynamic equilibrium branches respond to environmental changes.
The question is a natural one.
Matter concentrated in a finite radial region contributes an energy outside the horizon that depends on where the matter sits.
This energy can move the extrema of the free energy, and with them the number and stability of the equilibria.
 
Thermodynamic topology provides tools for studying this equilibrium structure~\cite{Wei:2022dzw,Wei:2021vdx,Wei:2026upy}.
In the generalized off-shell free-energy formulation, equilibria are zeros of a topological vector field.
Each isolated zero carries a topological charge, and the charges add up to a total topological number that classifies the BH~\cite{Wei:2022dzw,Fang:2022rsb,Wei:2024gfz}.
The method covers Gauss-Bonnet and Lovelock gravity, black strings, higher-dimensional and massive-gravity BHs, and charged dilaton BHs~\cite{Yerra:2022alz,Bai:2022klw,Hung:2023ggz,Malik:2023nso,Hazarika:2024xar}, with further extensions~\cite{Wu:2025xxo}.
It also tracks local changes: zeros with opposite charges can be created or annihilated in pairs, which leaves the total number unchanged~\cite{Wei:2022dzw,Ahmed:2022kyv}.
A fixed total number therefore does not imply a fixed equilibrium structure.
 
Thermodynamic topology has also been studied in matter backgrounds, including perfect-fluid dark matter~\cite{Rizwan:2023ivp}, quintessence~\cite{Zhang:2025cev}, and quantum-corrected BHs in Kiselev spacetimes~\cite{Sadeghi:2024moo}.
In some of these models, the environmental parameters change the range in which stable branches exist but keep the original topological class~\cite{Rizwan:2023ivp,Zhang:2025cev}.
These studies classify specific matter backgrounds.
They do not directly explain how radial rearrangements of a localized environment reorganize the equilibrium branches.
Answering this requires treating the environmental position as an independent control parameter and following the local defects as it varies.
We ask two questions.
Can the spatial configuration of a localized environment induce local topological reconstruction at fixed total topological number?
And how does this reconstruction depend on the BH background and on the environmental strength?
 
We study these questions on fixed, static, spherically symmetric backgrounds.
The environment is a phenomenological, radially localized energy distribution, and its peak position $a_m$ is the control parameter.
The profile and amplitude stay fixed, so varying $a_m$ isolates the effect of position.
Because the energy is counted outside the horizon, moving the peak also changes the total exterior environmental energy.
We add this energy to the generalized off-shell free energy at a prescribed bath temperature and construct the corresponding vector field.
The environment is a quasistatic configuration held by stresses or external constraints, and we neglect its backreaction and the Tolman redshift of the local temperature.
These simplifications isolate how an added environmental energy term shifts the extrema of the free energy.
 
Our main result is that, in the parameter ranges we consider, the position of a localized environment reconstructs the local defect structure and leaves the total topological number unchanged, provided no zero crosses the boundary of the physical domain.
In the Schwarzschild-AdS BH, our main example, shifting the environment creates or removes a stable-unstable pair of equilibria, which corresponds to the pair creation or annihilation of defects with charges $+1$ and $-1$.
We locate the critical positions from the degenerate-zero conditions and follow the reconstruction through free-energy landscapes, equilibrium branches, and parameter phase diagrams.
The environment thus changes the number and stability ordering of the branches and shifts the existing ones, changes that the total topological number cannot distinguish.
 
The same analysis for Schwarzschild, Reissner-Nordstr\"om (RN), and Reissner-Nordstr\"om-AdS (RN-AdS) BHs, which start from different topological classes, reveals the same mechanism.
The critical positions and the required environmental strength depend on the background.
In our examples, the required environmental energy ranges from below $1\%$ of the BH mass (Schwarzschild) to a value comparable to it (RN-AdS).
The RN-AdS case therefore lies in the strong-environment regime of the effective free-energy model.
Testing whether the reconstruction survives there requires a self-consistent treatment that includes gravitational backreaction and relativistic thermal equilibrium.
 
The paper is organized as follows.
Section~\ref{sec:framework} presents the environment model, the generalized off-shell free energy, and the topological vector field, and states the conditions for the quasistatic description.
Section~\ref{sec:topological-classification} defines the topological classification used to compare the backgrounds.
Section~\ref{sec:sads-reconstruction} determines the critical reconstruction conditions for the Schwarzschild-AdS BH and follows the defect evolution and parameter phase diagrams.
Section~\ref{sec:other-backgrounds} extends the analysis to the other three backgrounds and compares them.
Section~\ref{sec:conclusions} summarizes the results and discusses self-consistent environmental coupling and more general backgrounds.

\section{Model and theoretical framework}
\label{sec:framework}

This section builds an effective description of BH thermodynamics in a localized external environment.
The peak position $a_m$ of the environmental energy distribution enters the generalized off-shell free energy, whose stationary points and their stability we track as $a_m$ varies.
Representing these stationary points as zeros of a topological vector field gives a common language for locating the critical conditions for reconstruction and for comparing different BH backgrounds.

\subsection{Black hole background and localized matter environment}

Consider a static, spherically symmetric BH background,
\begin{equation}
    ds^2
    =
    -f(r)\,dt^2
    +\frac{dr^2}{g(r)}
    +r^2d\Omega^2,
    \label{eq:static-background}
\end{equation}
where $r$ is the areal radius and $r_h$ is the event horizon radius.
We use natural units, $G=c=\hbar=k_{\rm B}=1$.
All quantities are made dimensionless according to their dimensions, using the fixed width of the environmental profile as the reference length; dimensionless markers are omitted below.
For the Schwarzschild-AdS (SAdS) BH,
\begin{equation}
    f(r)=g(r)
    =
    1-\frac{2M}{r}+\frac{r^2}{l^2},
    \label{eq:sads-metric}
\end{equation}
where $l$ is the AdS curvature radius.
The horizon condition gives
\begin{equation}
    M_0(r_h)
    =
    \frac{r_h}{2}+\frac{r_h^3}{2l^2}.
    \label{eq:sads-mass}
\end{equation}
For the other BH backgrounds treated below, only the background thermodynamic quantities change; the environmental profile stays the same.

We model the environment by a localized radial energy distribution,
\begin{equation}
    \frac{dE_{\rm env}}{dr}
    =
    \epsilon_m\operatorname{sech}^2(r-a_m),
    \qquad r\geq r_h,
    \label{eq:matter density}
\end{equation}
where $\epsilon_m>0$ sets the effective amplitude per unit radial interval and the width is fixed to unity.
The peak position $a_m$ is then the only parameter that varies; changing it moves the environment without changing its width or amplitude.
When $a_m\leq r_h$, only the decaying tail lies outside the horizon, and $a_m$ still marks the position of the distribution.

More general density profiles are available for self-consistent models: numerical studies have coupled BHs to anisotropic fluids using Hernquist, NFW, and Einasto profiles~\cite{Figueiredo:2023gas}, and to localized dark-matter distributions in studies of quasinormal modes~\cite{Lan:2025brn}.
We instead use Eq.~\eqref{eq:matter density} to isolate how the radial position of a fixed profile affects the free-energy structure.
Here $dE_{\rm env}/dr$ is a distribution of effective energy, not the local energy density per unit volume that a static observer would measure.

Integrating outside the horizon gives the environmental energy,
\begin{equation}
    E_{\rm env}(r_h,a_m)
    =
    \int_{r_h}^{\infty}
    \frac{dE_{\rm env}}{dr}\,dr
    =
    \epsilon_m
    \left[1+\tanh(a_m-r_h)\right],
    \label{eq:environment-energy}
\end{equation}
which at fixed $\epsilon_m$ depends only on the relative position $\Delta a=a_m-r_h$.
Because the lower limit of integration moves with the horizon radius, a fixed amplitude does not mean a fixed exterior energy: this is how the environmental position enters the stationary-point condition for the free energy.
The full equilibrium structure depends on the background mass and entropy as well, so the reconstruction threshold cannot be read off from $\Delta a$ alone.

We treat the environment as a quasistatic configuration held in place by stresses or external constraints, and let it enter the effective thermodynamic potential only through Eq.~\eqref{eq:environment-energy}.
We do not solve the Einstein equations with an environmental stress-energy tensor; the fixed-background approximation neglects the environmental correction to the geometry.
Comparing different values of $r_h$ still moves along the family of background solutions, so state variables such as the mass vary with $r_h$ as usual.

\subsection{Generalized off-shell free energy and effective equilibrium}

Let $\tau>0$ be the inverse-temperature control parameter of the external heat bath, following the BH canonical ensemble and the thermodynamics of finite boundaries~\cite{York:1986it,Brown:1994sn}.
We define the generalized off-shell free energy as
\begin{equation}
    \mathcal{F}(r_h,\tau;a_m)
    =
    M(r_h)
    +E_{\rm env}(r_h,a_m)
    -\frac{S(r_h)}{\tau},
    \label{eq:generalized-free-energy}
\end{equation}
where $M$ and $S$ are the background's mass and entropy, and the environment enters only as an added energy term; we do not separately model the environmental entropy or the work needed to maintain the external constraints.
Equation~\eqref{eq:generalized-free-energy} is the effective potential used throughout this paper.

For the four-dimensional BHs in Einstein gravity considered here, the entropy obeys the area law~\cite{Bekenstein:1973ur},
\begin{equation}
    S(r_h)=\pi r_h^2.
    \label{eq:area-entropy}
\end{equation}
Together with Eq.~\eqref{eq:sads-mass}, this gives the free energy in the Schwarzschild-AdS background:
\begin{equation}
    \mathcal{F}
    =
    \frac{r_h}{2}
    +\frac{r_h^3}{2l^2}
    +\epsilon_m\left[1+\tanh(a_m-r_h)\right]
    -\frac{\pi r_h^2}{\tau}.
    \label{eq:sads-effective-free-energy}
\end{equation}

The off-shell description lets $r_h$ vary independently within the physically admissible domain without imposing the stationary-point condition in advance.
At fixed $\tau$, $a_m$, $\epsilon_m$, and the other background control parameters, the effective equilibrium branches satisfy
\begin{equation}
    \mathcal{F}_{,r_h}
    =
    M'(r_h)
    -\frac{S'(r_h)}{\tau}
    -\epsilon_m\operatorname{sech}^2(a_m-r_h)
    =
    0.
    \label{eq:effective-equilibrium}
\end{equation}
Hereafter, comma subscripts denote partial derivatives with respect to the indicated variable, and primes denote derivatives with respect to $r_h$ at fixed values of the other control parameters.
At a stationary point, $\mathcal{F}_{,r_hr_h}>0$ corresponds to a local minimum of the free energy and $\mathcal{F}_{,r_hr_h}<0$ to a local maximum.
Local stability in this paper refers only to variations along the $r_h$ direction in this effective potential.

The environmental corrections to the first and second derivatives of the free energy are, respectively,
\begin{align}
    \frac{\partial E_{\rm env}}{\partial r_h}
    &=
    -\epsilon_m\operatorname{sech}^2(a_m-r_h),
    \label{eq:environment-first-derivative}\\
    \frac{\partial^2 E_{\rm env}}{\partial r_h^2}
    &=
    -2\epsilon_m\operatorname{sech}^2(a_m-r_h)
    \tanh(a_m-r_h).
    \label{eq:environment-second-derivative}
\end{align}
Varying $a_m$ shifts the radial region where these localized corrections act, changing the slope and curvature of the free energy.
This is how the environmental position drives the migration of stationary points and, with it, local reconstruction.

To compare the environmental energy with the BH mass, we define
\begin{equation}
    \eta(r_h,a_m)
    =
    \frac{E_{\rm env}(r_h,a_m)}{M(r_h)}.
    \label{eq:eta-def}
\end{equation}
$\eta\ll1$ means the environmental energy is much smaller than the background BH mass; $\eta=\mathcal{O}(1)$ means the two are comparable.
This ratio measures relative energy, but it does not by itself determine the environment's contribution to the free-energy derivatives: even at small $\eta$, the environmental correction can alter the stationary-point structure wherever the original slope or curvature is small.
For each model below, we report $\eta$ together with the reconstruction conditions.
In the strong-environment regime, our results characterize the effective potential; their physical realization requires assessing the effects of geometric backreaction.

\subsection{Temperature parameter and quasistatic description}

Strict thermal equilibrium in a static gravitational field requires the local temperature to satisfy the Tolman relation~\cite{Tolman:1930ona},
\begin{equation}
    T_{\rm loc}(r)\sqrt{f(r)}
    =
    T_0,
    \label{eq:tolman-relation}
\end{equation}
where $T_0$ depends on the normalization of the static time coordinate, and $T_{\rm loc}(r)$ is the temperature measured by a local observer in the static region.
If the local heat-bath temperature is specified at radius $R$, one must use $T_{\rm loc}(R)=T_0/\sqrt{f(R)}$ and choose the boundary energy and thermodynamic variables consistently~\cite{Brown:1994gs}.
A finite-radius boundary is one way to impose heat-bath conditions, but it is not required for the Tolman relation to hold.

$\tau$ enters Eq.~\eqref{eq:generalized-free-energy} as an external control parameter, not as a quantity derived from the local thermodynamics of matter — a distinction visible in how the stationary-point temperature relates to the background Hawking temperature.
For the chosen backgrounds, the first law gives $M'=T_HS'$ when the other parameters, such as the charge and the AdS curvature radius, are fixed, so Eq.~\eqref{eq:effective-equilibrium} yields
\begin{equation}
    \frac{1}{\tau}
    =
    T_H(r_h)
    -
    \frac{\epsilon_m
    \operatorname{sech}^2(a_m-r_h)}{S'(r_h)}.
    \label{eq:effective-temperature}
\end{equation}
When the environmental term is nonzero, this differs from $1/\tau=T_H$.
The branches below are therefore branches of stationary points of the effective free energy, not self-consistent thermal-equilibrium solutions of the background BH and its matter environment; such a description would need a consistent treatment of the environmental energy, entropy, constraints, and temperature redshift.

Varying $a_m$ compares a family of static effective potentials with different external configurations.
If this variation stands in for an actual slow control process, its timescale must exceed the relaxation time near a stable branch, so that the system can approximately follow neighboring local equilibria~\cite{Yang:2021KerrLandscape}.
Unstable branches describe the free-energy structure, but they are not states in which the system can remain for long, and quasistatic following can also fail near branch degeneracies.
We compute the branches and their bifurcations without determining the actual transition processes or rates — a dynamical question distinct from the effect of infalling matter on BH waveforms~\cite{Tian:2025uvk}.

\subsection{Topological vector field and topological number}

Following the construction of BH thermodynamic topology~\cite{Wei:2021vdx,Wei:2022dzw}, we define, at fixed external control parameters,
\begin{equation}
    \boldsymbol{\phi}
    =
    \left(
        \mathcal{F}_{,r_h},
        -\cot\theta\,\csc\theta
    \right),
    \qquad 0<\theta<\pi,
    \label{eq:topological-vector}
\end{equation}
where $r_h$ lies in the physically admissible domain of the chosen BH.
The auxiliary parameter $\theta$ is not an angular coordinate of the spacetime metric.
The zeros of the vector field satisfy
\begin{equation}
    \mathcal{F}_{,r_h}=0,
    \qquad
    \theta=\frac{\pi}{2},
    \label{eq:topological-zero}
\end{equation}
and are therefore in one-to-one correspondence with the stationary points of the effective free energy.

Consider a closed contour $\partial\Sigma_i$ enclosing only the $i$th isolated zero, with $\boldsymbol{\phi}\neq0$ everywhere on the contour and traversed in the positive orientation of the $(r_h,\theta)$ plane.
The topological charge of this zero is defined as
\begin{equation}
    w_i
    =
    \frac{1}{2\pi}
    \oint_{\partial\Sigma_i}d\Omega,
    \qquad
    \Omega
    =
    \arg\!\left(\phi^{r_h}+{\rm i}\phi^\theta\right),
    \label{eq:winding-definition}
\end{equation}
where the phase is continuously unwrapped along the integration path.
The sum of the charges of all isolated zeros in the region under consideration is
\begin{equation}
    W=\sum_i w_i.
    \label{eq:total-topological-number}
\end{equation}

Since $\phi^{r_h}$ is independent of $\theta$ and $\phi^\theta$ is independent of $r_h$, the Jacobian matrix at a zero $(r_i,\pi/2)$ is
\begin{equation}
    \left.
    \frac{\partial(\phi^{r_h},\phi^\theta)}
         {\partial(r_h,\theta)}
    \right|_{(r_i,\pi/2)}
    =
    \begin{pmatrix}
        \mathcal{F}_{,r_hr_h}(r_i) & 0\\
        0 & 1
    \end{pmatrix}.
    \label{eq:zero-jacobian}
\end{equation}
For a nondegenerate zero, $\mathcal{F}_{,r_hr_h}(r_i)\neq0$, and the local topological charge equals the sign of the Jacobian determinant:
\begin{equation}
    w_i
    =
    \operatorname{sign}
    \left[\mathcal{F}_{,r_hr_h}(r_i)\right].
    \label{eq:wi-analytic}
\end{equation}
With our convention, $w_i=+1$ corresponds to a local minimum along $r_h$, and $w_i=-1$ to a local maximum.
Equation~\eqref{eq:wi-analytic} applies to all backgrounds using the construction in Eq.~\eqref{eq:topological-vector}, but not to degenerate zeros, where the Jacobian determinant vanishes.

Suppose $\mathcal{F}_{,r_h}$ is continuous and two neighboring zeros are simple, with no other zeros between them.
The function keeps the same sign between the zeros and so crosses zero in opposite directions at the two endpoints, which means the charges of neighboring simple zeros alternate in sign.
As $a_m$ varies, a generic interior pair creation or annihilation event occurs at
\begin{equation}
    \mathcal{F}_{,r_h}(r_c,a_c)=0,
    \qquad
    \mathcal{F}_{,r_hr_h}(r_c,a_c)=0.
    \label{eq:degenerate-zero}
\end{equation}
If, in addition,
\begin{equation}
    \mathcal{F}_{,r_hr_hr_h}(r_c,a_c)\neq0,
    \qquad
    \mathcal{F}_{,r_ha_m}(r_c,a_c)\neq0,
    \label{eq:fold-nondegeneracy}
\end{equation}
the local expansion takes the form of a fold bifurcation:
\begin{equation}
    \phi^{r_h}
    \simeq
    \mathcal{F}_{,r_ha_m}(r_c,a_c)(a_m-a_c)
    +
    \frac{1}{2}
    \mathcal{F}_{,r_hr_hr_h}(r_c,a_c)(r_h-r_c)^2.
    \label{eq:fold-expansion}
\end{equation}
Two simple zeros exist on one side of the critical point and the pair is absent on the other; their charges are $+1$ and $-1$, so their total local charge vanishes.
This is related to vortex-antivortex pair creation in the thermodynamic free-energy description~\cite{Ahmed:2022kyv}.
Higher-order degeneracies and events at the boundary of parameter space require separate analysis.

To cross-check the numerical implementation, we compute contour winding numbers for representative zeros using Eq.~\eqref{eq:winding-definition} and compare them with Eq.~\eqref{eq:wi-analytic}.
The results agree, as Table~\ref{tab:contour-check} shows.
This comparison tests the numerical implementation and sign convention; the analytical criterion itself follows from Eq.~\eqref{eq:zero-jacobian}.

\begin{table}[htbp!]
    \centering
    \captionsetup{width=0.9\textwidth}
    \caption{
    Comparison of the analytical charge criterion and numerical contour winding numbers for nondegenerate zeros.
    The SAdS parameters are $a_m=4$, $\tau=3.25$, $\epsilon_m=3.5$, and $l=2$.
    The RN parameters are $a_m=3.7$, $\tau=80$, $\epsilon_m=0.5$, and $Q=1$.
    Zero locations are displayed to six decimal places; the contour calculations use unrounded values.
    }
    \label{tab:contour-check}
    \begin{tabular}{@{}lccc@{}}
        \toprule
        Model & Zero $r_i$ & Analytical $w_i$ & Contour winding number\\
        \midrule
        SAdS & 0.268455 & $-1$ & $-1.000000$\\
        SAdS & 5.331833 & $+1$ & $+1.000000$\\
        RN   & 1.116604 & $+1$ & $+1.000000$\\
        RN   & 2.726583 & $-1$ & $-1.000000$\\
        RN   & 5.511301 & $+1$ & $+1.000000$\\
        RN   & 5.826928 & $-1$ & $-1.000000$\\
        \bottomrule
    \end{tabular}
\end{table}

Conservation of the total topological number also depends on the boundary of the parameter domain.
Suppose that, as $a_m$ varies continuously, a fixed bounded region can be chosen to contain all zeros under consideration, with the vector field continuous and nonzero on its boundary throughout; the winding number of the boundary map is then invariant, and $W$ is conserved.
Interior defects can still migrate or undergo pair creation or annihilation, changing the number of equilibrium branches and their local stability structure.
This conservation argument does not directly apply if a zero crosses the physical boundary or if the boundary conditions change during the evolution.
For each model below, we check both the local degeneracy conditions and the physical boundary to distinguish local topological reconstruction from changes in the total topological number.

\section{Thermodynamic topological classification of black holes}
\label{sec:topological-classification}

Section~\ref{sec:framework} relates stationary points of the effective free energy to zeros of the topological vector field; thermodynamic topological classification then uses the total topological number and the stability ordering of equilibrium branches to compare the global structure of BH systems~\cite{Wei:2024gfz,Zhu:2024RotatingClasses}.
This section states the classification criteria used below and distinguishes the global topological class from the local defect reconstruction that the environment induces.

For the vector field constructed in Sec.~\ref{sec:framework}, Eq.~\eqref{eq:wi-analytic} gives the charge of each nondegenerate zero: $w_i=+1$ for a local minimum of the effective free energy along $r_h$, and $w_i=-1$ for a local maximum.
A single nondegenerate zero can carry only $\pm1$, so a total $W=0$ can come from cancellation between opposite charges — it does not imply the absence of equilibrium branches.

At fixed external control parameters, order all simple zeros in the physically admissible domain by increasing horizon radius:
\begin{equation}
    r_1<r_2<\cdots<r_N.
    \label{eq:ordered-equilibrium-roots}
\end{equation}
For the one-dimensional stationary-point structure used here, neighboring simple zeros have alternating charges, so for $N\geq1$ their sum is
\begin{equation}
    W=\sum_{i=1}^{N}w_i
    =
    \begin{cases}
        w_1, & N \text{ is odd},\\
        0,   & N \text{ is even}.
    \end{cases}
    \label{eq:classification-charge-sum}
\end{equation}
$W$ therefore distinguishes total charges of $-1$, $0$, and $+1$, but not the two stability orderings possible when $W=0$, since the smallest-radius branch can be either stable or unstable.

Including the stability of the smallest- and largest-radius branches gives four classes, $W^{1-}$, $W^{0+}$, $W^{0-}$, and $W^{1+}$~\cite{Wei:2024gfz,Zhu:2024RotatingClasses}, where the digit denotes $|W|$ and the sign denotes the charge of the smallest-radius branch.
Table~\ref{tab:thermodynamic-topological-classes} lists these classes and the backgrounds, without an environment, used in this study.
Here small- and large-radius branches mean the smallest and largest equilibrium radii available at a given temperature, not predefined radius intervals.

\begin{table}[htbp!]
    \centering
    \captionsetup{width=0.9\textwidth}
    \caption{
    Four thermodynamic topological classes and the corresponding backgrounds without an environment used in this study.
    Stability refers to local stability of the relevant free energy along the horizon-radius direction.
    Charged backgrounds have fixed nonzero charge, and AdS backgrounds have fixed curvature radius.
    Branch stabilities are compared within temperature ranges admitting nondegenerate equilibria.
    }
    \label{tab:thermodynamic-topological-classes}
    \begin{tabular}{@{}ccccc@{}}
        \toprule
        Class
        & $W$
        & \shortstack{Smallest-radius\\branch}
        & \shortstack{Largest-radius\\branch}
        & \shortstack{Background without\\an environment}\\
        \midrule
        $W^{1-}$
        & $-1$
        & Unstable
        & Unstable
        & Schwarzschild\\
        $W^{0+}$
        & $0$
        & Stable
        & Unstable
        & RN\\
        $W^{0-}$
        & $0$
        & Unstable
        & Stable
        & Schwarzschild-AdS\\
        $W^{1+}$
        & $+1$
        & Stable
        & Stable
        & RN-AdS\\
        \bottomrule
    \end{tabular}
\end{table}

When no equilibrium zeros exist in some temperature interval, $W$ is zero as an empty sum, which alone does not distinguish $W^{0+}$ from $W^{0-}$: the class must then be read from the stability ordering where branches do exist, or from the asymptotic behavior of the vector field at the physical boundary.
Table~\ref{tab:thermodynamic-topological-classes} gives a baseline for the comparisons that follow; once the environment is introduced, the zeros and boundary behavior of the effective potential must still be checked directly.

A global topological class does not capture all local equilibrium information.
The radially ordered charge sequences
\begin{equation}
    (-1,+1)
    \qquad\text{and}\qquad
    (-1,+1,-1,+1)
    \label{eq:equal-class-distinct-branches}
\end{equation}
both give $W=0$ and the same stability at the two endpoint branches, yet the second sequence contains one additional stable-unstable pair.
If the boundary map of the region stays continuous and nonzero, interior defects can be created or annihilated in pairs without changing the total topological number — local topological reconstruction can occur within a single global class.
A zero that crosses the physical boundary, however, forces the total topological number and the classification to be reassessed.

We track the number, radial positions, and charge ordering of zeros to identify the local changes that the environmental configuration induces.
The next section takes up the Schwarzschild-AdS BH first: without an environment it belongs to $W^{0-}$, and we determine the critical conditions under which the environment creates or removes additional equilibrium branches.
We then apply the same method to the other three backgrounds to compare their environmental responses.

\section{Topological reconstruction in the Schwarzschild-AdS black hole}
\label{sec:sads-reconstruction}

The Schwarzschild-AdS BH admits thermodynamic branches with different stabilities and has long been a testing ground for BH free-energy structure and phase transitions~\cite{Hawking:1982dh,Wei:2020HP,Yerra:2022coh,Hazarika:2024imk,Prestidge:1999uq}.
Using the topological description of BH thermodynamics~\cite{Wei:2026upy} and the effective free-energy model of Sec.~\ref{sec:framework}, we ask how the radial position of a localized environment changes the number, location, and local stability of these branches, with $a_m$ as the control parameter for interior degenerate zeros.
Free-energy landscapes, topological vector fields, and equilibrium branch diagrams together describe the resulting local reconstruction at fixed total topological number.

\subsection{Effective free energy and equilibrium branches}

For the Schwarzschild-AdS background introduced in Sec.~\ref{sec:framework}, the mass, entropy, and Hawking temperature are
\begin{equation}
    M(r_h)
    =
    \frac{r_h}{2}+\frac{r_h^3}{2l^2},
    \qquad
    S(r_h)=\pi r_h^2,
    \qquad
    T_H(r_h)
    =
    \frac{l^2+3r_h^2}{4\pi l^2r_h},
    \label{eq:sads-thermodynamic-data}
\end{equation}
where $T_H$ is determined by the surface gravity at the horizon~\cite{Hawking:1975vcx}.
We retain the dimensionless convention of Sec.~\ref{sec:framework} and fix $l$ and $\tau$ throughout the following analysis.

Including the environmental energy outside the horizon gives the generalized off-shell free energy
\begin{equation}
    \mathcal{F}(r_h,\tau;a_m)
    =
    \frac{r_h}{2}
    +\frac{r_h^3}{2l^2}
    -\frac{\pi r_h^2}{\tau}
    +\epsilon_m\left[1+\tanh(a_m-r_h)\right],
    \label{eq:sads-free-energy}
\end{equation}
whose radial derivative we denote by
\begin{equation}
    G(r_h;a_m,\epsilon_m)
    \equiv
    \frac{\partial\mathcal{F}}{\partial r_h}
    =
    \frac{1}{2}
    +\frac{3r_h^2}{2l^2}
    -\frac{2\pi r_h}{\tau}
    -\epsilon_m\operatorname{sech}^2(a_m-r_h).
    \label{eq:sads-G}
\end{equation}
The corresponding topological vector field, restricted to $r_h>0$, is
\begin{equation}
    \boldsymbol{\phi}
    =
    \left(
        G,\,
        -\cot\theta\,\csc\theta
    \right),
    \label{eq:sads-vector}
\end{equation}
and its zeros satisfy
\begin{equation}
    G=0,
    \qquad
    \theta=\frac{\pi}{2}.
    \label{eq:sads-zero}
\end{equation}
The equilibrium branches associated with these zeros, and their stability, follow the effective potential of Sec.~\ref{sec:framework}.

The environmental correction to the radial component,
\begin{equation}
    \Delta G
    =
    -\epsilon_m\operatorname{sech}^2(a_m-r_h)
    \label{eq:sads-environment-correction}
\end{equation}
is localized near $r_h\simeq a_m$.
Varying $a_m$ shifts this localized negative correction across different radial regions of the background free energy; where it changes the local zero-crossing structure of $G$, a pair of neighboring free-energy extrema appears or disappears.

\subsection{Critical positions for interior reconstruction}

At fixed $\epsilon_m$, degenerate points of interior equilibrium branches satisfy
\begin{equation}
    G(r_h;a_m,\epsilon_m)=0,
    \qquad
    G_{,r_h}(r_h;a_m,\epsilon_m)=0,
    \label{eq:sads-fold-conditions}
\end{equation}
where
\begin{equation}
    G_{,r_h}
    =
    \frac{3r_h}{l^2}
    -\frac{2\pi}{\tau}
    -2\epsilon_m
    \operatorname{sech}^2(a_m-r_h)
    \tanh(a_m-r_h).
    \label{eq:sads-G-derivative}
\end{equation}
If $G_{,r_hr_h}\neq0$ and $G_{,a_m}\neq0$ at a degenerate point, it is a fold bifurcation with $a_m$ as the control parameter, and the number of simple zeros on its two sides differs by two.

We choose
\begin{equation}
    l=2,
    \qquad
    \tau=3.25,
    \qquad
    \epsilon_m=3.5.
    \label{eq:sads-reference-parameters}
\end{equation}
Numerically solving Eq.~\eqref{eq:sads-fold-conditions} yields
\begin{align}
    \left(r_h^{(1)},a_m^{(1)}\right)
    &\simeq
    (5.644203,\,6.615737),
    \label{eq:sads-first-fold}\\
    \left(r_h^{(2)},a_m^{(2)}\right)
    &\simeq
    (6.083252,\,6.635772).
    \label{eq:sads-second-fold}
\end{align}
Both points satisfy the nondegeneracy conditions above and correspond, respectively, to local defect pair creation and annihilation; the additional equilibrium branches exist within the environmental-position window
\begin{equation}
    a_m^{(1)}<a_m<a_m^{(2)},
    \qquad
    a_m^{(2)}-a_m^{(1)}
    \simeq 0.020035.
    \label{eq:sads-reconstruction-window}
\end{equation}

Interior reconstruction must also be distinguished from changes at the physical boundary.
Equation~\eqref{eq:sads-G} gives
\begin{equation}
    G(0^+)
    =
    \frac{1}{2}
    -\epsilon_m\operatorname{sech}^2a_m,
    \qquad
    G(r_h)\longrightarrow+\infty
    \quad(r_h\longrightarrow\infty).
    \label{eq:sads-boundary-signs}
\end{equation}
For $\epsilon_m=3.5$ and $a_m\geq0$, the sign at the lower boundary changes at
\begin{equation}
    a_{\rm b}
    =
    \operatorname{arcosh}\sqrt{2\epsilon_m}
    \simeq 1.628307.
    \label{eq:sads-boundary-position}
\end{equation}
Both interior critical points considered here lie in $a_m>a_{\rm b}$, where $G$ is positive at both the small- and large-radius boundaries, so interior defects can be created or annihilated in pairs while $W=0$ is preserved.
The three representative configurations below are all chosen from this range.

\subsection{Free energy landscapes and topological defect evolution}

\begin{figure}[htbp!]
    \centering
    \includegraphics[width=0.7\textwidth]{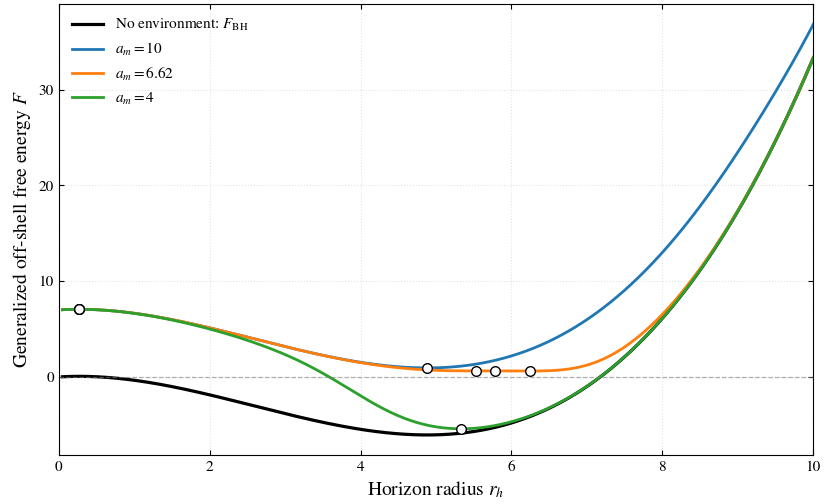}
    \captionsetup{width=0.9\textwidth}
    \caption{
    Generalized off-shell free energy landscapes of the
    Schwarzschild-AdS BH.
    The parameters are $l=2$, $\epsilon_m=3.5$, and $\tau=3.25$.
    The curves with $a_m=4$, $6.62$, and $10$ represent
    configurations before, within, and after the interval
    of interior defect reconstruction, respectively.
    The marked extrema satisfy
    $\partial\mathcal{F}/\partial r_h=0$ and represent
    stationary states of the effective model.
    }
    \label{fig:free_energy_landscape}
\end{figure}

We choose
\begin{equation}
    a_m=4,\qquad 6.62,\qquad 10,
    \label{eq:sads-representative-positions}
\end{equation}
to represent the equilibrium structures before the first interior reconstruction, between the two critical points, and after the second reconstruction, respectively.
Figure~\ref{fig:free_energy_landscape} shows the corresponding free energy, whose extrema correspond one-to-one to the zeros in Eq.~\eqref{eq:sads-zero}, so their number and type directly characterize the local defect structure.

At $a_m=4$, the free energy has one local maximum and one local minimum.
When $a_m$ enters the window in Eq.~\eqref{eq:sads-reconstruction-window}, the number of extrema increases to four: a maximum, minimum, maximum, and minimum, in order of increasing radius.
Beyond the second critical position, only two extrema remain.
This change is the pair creation or annihilation of locally stable and unstable branches in the effective model; it does not by itself determine the transition path the system follows.

\begin{figure}[htbp!]
    \centering
    \begin{subfigure}[t]{0.31\textwidth}
        \centering
        \includegraphics[width=\linewidth]{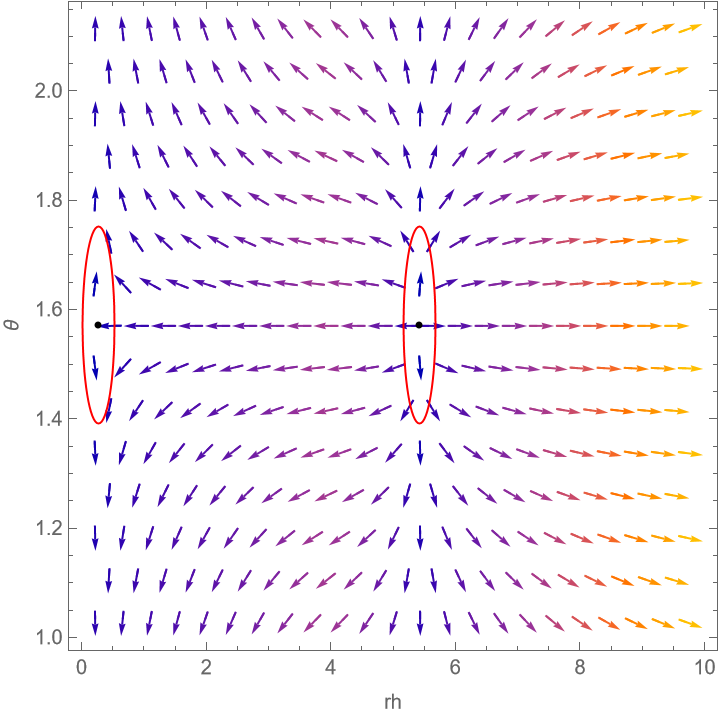}
        \caption{$a_{\rm b}<a_m<a_m^{(1)}$}
        \label{fig:swz_region1}
    \end{subfigure}
    \hfill
    \begin{subfigure}[t]{0.31\textwidth}
        \centering
        \includegraphics[width=\linewidth]{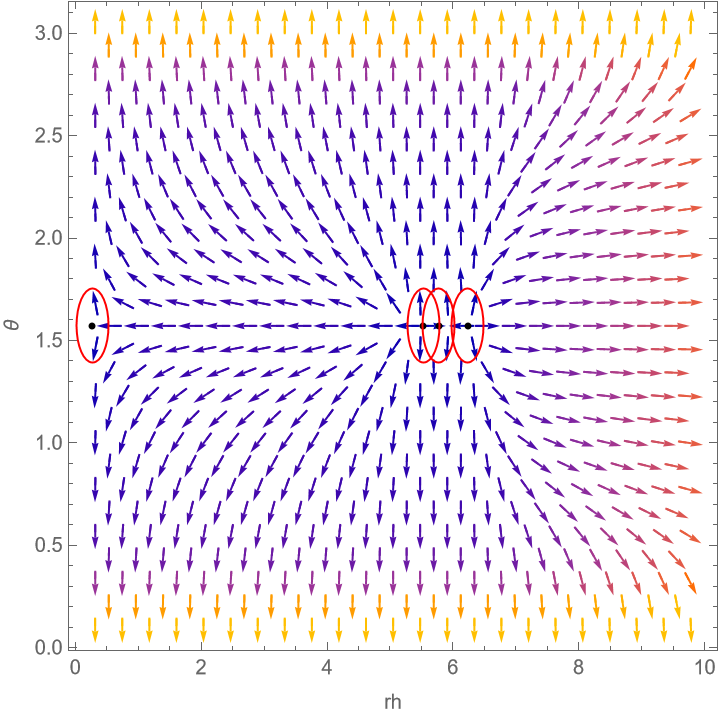}
        \caption{$a_m^{(1)}<a_m<a_m^{(2)}$}
        \label{fig:swz_region2}
    \end{subfigure}
    \hfill
    \begin{subfigure}[t]{0.31\textwidth}
        \centering
        \includegraphics[width=\linewidth]{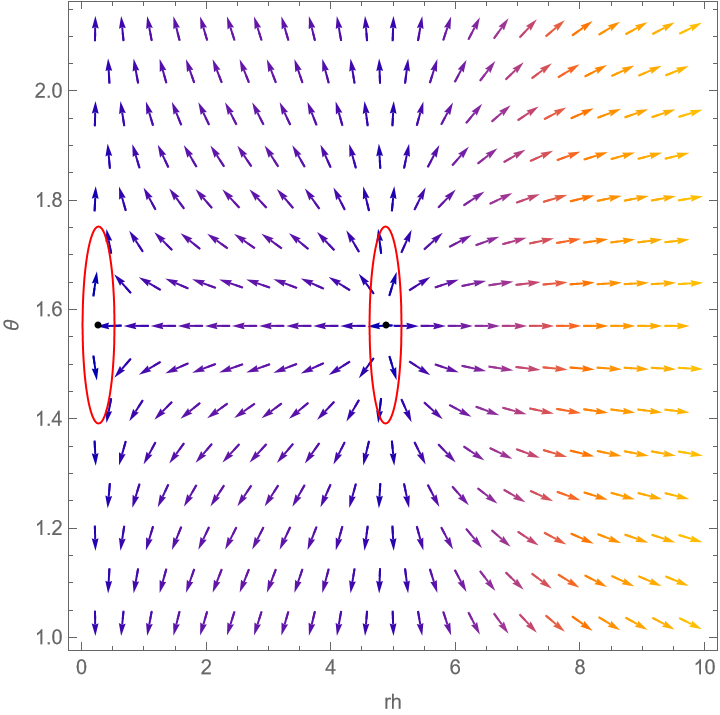}
        \caption{$a_m>a_m^{(2)}$}
        \label{fig:swz_region3}
    \end{subfigure}
    \captionsetup{width=0.9\textwidth}
    \caption{
    Topological vector fields of the Schwarzschild-AdS BH at three representative environmental positions.
    The parameters are $l=2$, $\epsilon_m=3.5$, and $\tau=3.25$.
    The left, middle, and right panels correspond to $a_m=4$, $6.62$, and $10$, respectively.
    }
    \label{fig:swzads}
\end{figure}

Figure~\ref{fig:swzads} shows the topological vector field for the same parameters, with all zeros lying at $\theta=\pi/2$; their radial positions and charges are listed in Table~\ref{tab:sads_charges}.

\begin{table}[htbp!]
    \centering
    \captionsetup{width=0.9\textwidth}
    \caption{
    Zero locations, local topological charges, and total topological number for the Schwarzschild-AdS effective model.
    The parameters are $l=2$, $\epsilon_m=3.5$, and $\tau=3.25$.
    Zeros are ordered by increasing radius, with locations rounded to three decimal places.
    Charges are determined from Eq.~\eqref{eq:wi-analytic}.
    }
    \label{tab:sads_charges}
    \begin{tabular}{@{}lccc@{}}
        \toprule
        $a_m$ & Zeros $r_i$ & Charges $(w_i)$ & $W$\\
        \midrule
        $4$
        & $0.268,\ 5.332$
        & $-1,\ +1$
        & $0$\\
        $6.62$
        & $0.273,\ 5.534,\ 5.779,\ 6.243$
        & $-1,\ +1,\ -1,\ +1$
        & $0$\\
        $10$
        & $0.273,\ 4.883$
        & $-1,\ +1$
        & $0$\\
        \bottomrule
    \end{tabular}
\end{table}

At $a_m^{(1)}$, a new pair of zeros appears between the two original ones: the smaller-radius zero carries $w=+1$ and marks a new locally stable branch, while the larger-radius zero carries $w=-1$ and marks a new unstable branch.
As $a_m$ increases, this new unstable branch approaches the original large-radius stable branch, and the two merge and annihilate at $a_m^{(2)}$; the new stable branch survives, and the original small-radius unstable branch persists throughout.
Both events involve opposite charges, but the branches annihilated in the second event are not the pair created in the first.

This evolution illustrates defect pair creation and annihilation in BH thermodynamic topology~\cite{Ahmed:2022kyv,Wei:2022dzw}.
Outside the interval between the critical points, the simple zeros carry the ordered charges $(-1,+1)$; inside the reconstruction window, the sequence becomes $(-1,+1,-1,+1)$, so the number of local zeros changes as $2\rightarrow4\rightarrow2$ while the total topological number stays zero.
The critical zeros are themselves degenerate, so their charges cannot be read off directly from the sign criterion for nondegenerate zeros.

\subsection{Equilibrium branch diagrams and dependence on environmental strength}

\begin{figure}[htbp!]
    \centering
    \includegraphics[width=0.62\linewidth]{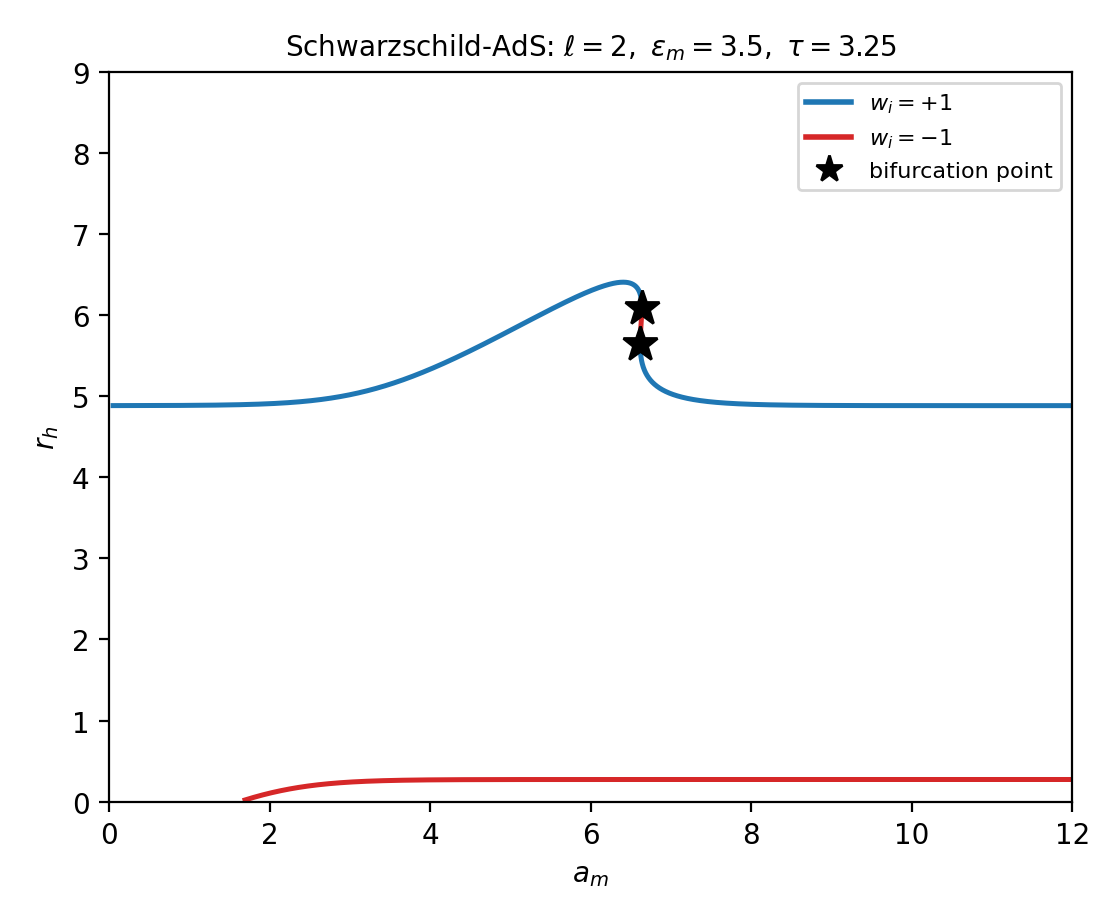}
    \captionsetup{width=0.9\textwidth}
    \caption{
    Equilibrium branches of the Schwarzschild-AdS effective model in the $(a_m,r_h)$ plane, with $l=2$, $\epsilon_m=3.5$, and $\tau=3.25$.
    Blue and red indicate nondegenerate branches with $w_i=+1$ and $w_i=-1$, respectively.
    Black stars mark the two interior fold points.
    The small-radius branch meets the $r_h=0$ boundary at $a_m=a_{\rm b}\simeq1.628307$.
    }
    \label{fig:sads_phase}
\end{figure}

To trace the zeros continuously, we rewrite Eq.~\eqref{eq:sads-G} as
\begin{equation}
    G
    =
    B(r_h)
    -\epsilon_m\operatorname{sech}^2(a_m-r_h),
    \qquad
    B(r_h)
    =
    \frac{1}{2}
    +\frac{3r_h^2}{2l^2}
    -\frac{2\pi r_h}{\tau}.
    \label{eq:sads-background-function}
\end{equation}
For $0<B(r_h)\leq\epsilon_m$, the equilibrium branches can be parametrized as
\begin{equation}
    a_m(r_h)
    =
    r_h
    \pm
    \operatorname{arcosh}
    \sqrt{\frac{\epsilon_m}{B(r_h)}}.
    \label{eq:sads-branch-parametrization}
\end{equation}
Retaining the portions with $r_h>0$ and $a_m$ in the chosen range gives the equilibrium branches in Fig.~\ref{fig:sads_phase}, each colored, where nondegenerate, according to $w_i=\operatorname{sign}(G_{,r_h})$.

The two interior fold points delimit the narrow window in which the additional zeros exist.
At fixed $a_m$, the number of intersections between a vertical line and the branches equals the number of equilibria: four between the folds and two outside them, provided $a_m>a_{\rm b}$.
At smaller $a_m$, the small-radius branch terminates at $r_h=0$, signaling a change at the physical boundary.

Away from degeneracies, the alternating signs of simple zeros give
\begin{equation}
    W
    =
    \frac{
    \operatorname{sign}G(\infty)
    -
    \operatorname{sign}G(0^+)
    }{2}.
    \label{eq:sads-W-boundary}
\end{equation}
For the fixed parameters in this section, $W=+1$ when $0\leq a_m<a_{\rm b}$, whereas $W=0$ when $a_m>a_{\rm b}$.
So the conservation of $W$ found for interior defect pair reconstruction does not extend to the full range of $a_m$ once boundary crossings are included.

\begin{figure}[htbp!]
    \centering
    \includegraphics[width=0.62\linewidth]{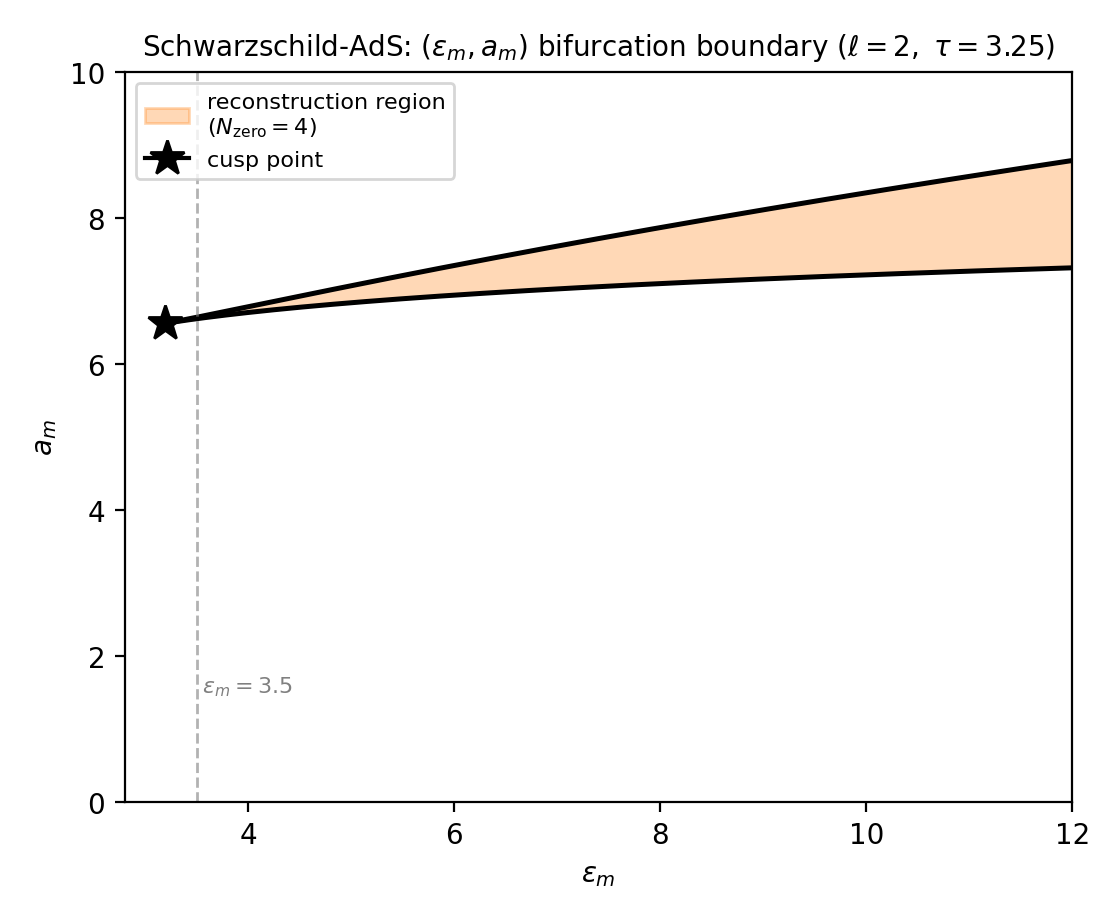}
    \captionsetup{width=0.9\textwidth}
    \caption{
    Region of interior topological reconstruction for the Schwarzschild-AdS effective model in the $(\epsilon_m,a_m)$ plane, with $l=2$ and $\tau=3.25$.
    The solid black curves satisfy $G=G_{,r_h}=0$ and meet at $(\epsilon_m^{\rm c},a_m^{\rm c})\simeq(3.188953,6.550229)$.
    The orange region contains four simple zeros.
    The gray dashed line denotes $\epsilon_m=3.5$.
    The boundary-crossing curve associated with $r_h=0$ is not shown.
    }
    \label{fig:sads_eps_scan}
\end{figure}

Varying $\epsilon_m$ determines the region in which interior reconstruction exists.
Along the fold boundary, Eq.~\eqref{eq:sads-fold-conditions} gives
\begin{equation}
    u(r_h)
    \equiv
    \tanh(a_m-r_h)
    =
    \frac{B'(r_h)}{2B(r_h)}.
    \label{eq:sads-fold-u}
\end{equation}
For $B(r_h)>0$ and $|u(r_h)|<1$, the critical curves can be parametrized as
\begin{equation}
    a_m
    =
    r_h+\operatorname{artanh}u(r_h),
    \qquad
    \epsilon_m
    =
    \frac{B(r_h)}{1-u(r_h)^2},
    \label{eq:sads-fold-parametrization}
\end{equation}
giving the two interior reconstruction boundaries shown in Fig.~\ref{fig:sads_eps_scan}.

The two boundaries meet at a cusp satisfying
\begin{equation}
    G=0,
    \qquad
    G_{,r_h}=0,
    \qquad
    G_{,r_hr_h}=0
    \label{eq:sads-cusp-conditions}
\end{equation}
where
\begin{equation}
    G_{,r_hr_h}
    =
    \frac{3}{l^2}
    +
    2\epsilon_m\operatorname{sech}^2(a_m-r_h)
    \left[1-3\tanh^2(a_m-r_h)\right].
    \label{eq:sads-G-second-derivative}
\end{equation}
For $l=2$ and $\tau=3.25$, the numerical result is
\begin{equation}
    \left(
        r_h^{\rm c},
        a_m^{\rm c},
        \epsilon_m^{\rm c}
    \right)
    \simeq
    (5.806706,\,6.550229,\,3.188953).
    \label{eq:sads-cusp}
\end{equation}
The third radial derivative at this point is nonzero and $(a_m,\epsilon_m)$ provide independent unfoldings of the degeneracy conditions, so the point satisfies the nondegeneracy requirements of a generic cusp bifurcation.

Near the cusp, this pair of interior folds is absent for $\epsilon_m<\epsilon_m^{\rm c}$ and, for $\epsilon_m>\epsilon_m^{\rm c}$, two critical boundaries emerge and enclose a region with four zeros; within the plotted range, the window of $a_m$ admitting interior reconstruction widens as $\epsilon_m$ increases.
Thus $\epsilon_m^{\rm c}$ is the amplitude threshold for this mechanism at fixed $l$, $\tau$, and environmental profile — not a universal constant across backgrounds or temperatures.

Outside the orange region, the number of zeros must also be determined from the physical boundary.
For $a_m\geq0$, the boundary-crossing condition is
\begin{equation}
    \epsilon_m
    =
    \frac{1}{2}\cosh^2a_m.
    \label{eq:sads-boundary-curve}
\end{equation}
The region outside the orange area therefore cannot be identified as a two-zero region; the plotted small-$a_m$ region, for instance, also contains configurations with one zero and $W=+1$.
The interior fold boundaries and Eq.~\eqref{eq:sads-boundary-curve} together give the complete partition by zero count.

For $\epsilon_m=3.5$, Eq.~\eqref{eq:eta-def} gives the energy ratios at the two interior critical points:
\begin{equation}
    \eta\left(r_h^{(1)},a_m^{(1)}\right)
    \simeq0.2420,
    \qquad
    \eta\left(r_h^{(2)},a_m^{(2)}\right)
    \simeq0.1686.
    \label{eq:sads-critical-energy-ratios}
\end{equation}
The environmental energy is smaller than the background BH mass at both points, but neither ratio establishes a clear hierarchy $\eta\ll1$, so these values alone do not justify neglecting geometric backreaction; they instead quantify the environmental energy this example requires for reconstruction.
The physical validity of the effective fixed-background model still depends on a more complete description of environmental coupling.

\section{Extensions to black holes in different topological classes}
\label{sec:other-backgrounds}

The thermodynamic topological classification of BH backgrounds gives a basis for comparing their environmental responses~\cite{Fang:2022rsb,Wei:2021vdx,Wei:2024gfz,Wang:2024RegularTopology}.
We extend the analysis to Schwarzschild, Reissner-Nordstr\"om (RN), and Reissner-Nordstr\"om-AdS (RN-AdS) BHs, asking whether the same localized environmental profile can reconstruct equilibrium branches in different initial topological classes, and how the critical conditions depend on the background.

All three models use the environmental energy and area entropy defined in Sec.~\ref{sec:framework}.
Their generalized off-shell free energy takes the common form
\begin{equation}
    \mathcal{F}
    =
    M(r_h)
    -\frac{\pi r_h^2}{\tau}
    +\epsilon_m\left[1+\tanh(a_m-r_h)\right].
    \label{eq:other-backgrounds-free-energy}
\end{equation}
The radial component of the topological vector field is
\begin{equation}
    G(r_h;a_m,\tau)
    \equiv
    \phi^{r_h}
    =
    M'(r_h)
    -\frac{2\pi r_h}{\tau}
    -\epsilon_m\operatorname{sech}^2(a_m-r_h).
    \label{eq:master-G}
\end{equation}
Each background enters only through the mass function $M(r_h)$ and the physically admissible domain of the horizon radius; the environmental term keeps the same form, and parameters such as the charge, the AdS curvature radius, and the environmental amplitude are held fixed when taking derivatives.
We keep the dimensionless convention and take $a_m>0$ below.

Equilibrium branches satisfy $G=0$, and interior fold points obey
\begin{equation}
    G=0,
    \qquad
    G_{,r_h}=0,
    \label{eq:other-backgrounds-fold}
\end{equation}
with $G_{,r_hr_h}\neq0$ and $G_{,a_m}\neq0$ — saddle-node bifurcations describing the creation or annihilation of neighboring equilibrium branches~\cite{Hazarika:2025esd}.
The charge of a nondegenerate branch is still given by $w_i=\operatorname{sign}(G_{,r_h})$, and for each background below we check both the interior degeneracy conditions and the physical boundary to distinguish local reconstruction from changes in the total topological number.

\subsection{Topological structure of the Schwarzschild BH}

The mass and Hawking temperature of the Schwarzschild BH are
\begin{equation}
    M(r_h)=\frac{r_h}{2},
    \qquad
    T_H(r_h)=\frac{1}{4\pi r_h},
    \qquad r_h>0.
    \label{eq:schw-background-data}
\end{equation}
Substitution into Eq.~\eqref{eq:master-G} gives
\begin{equation}
    G
    =
    \frac{1}{2}
    -\frac{2\pi r_h}{\tau}
    -\epsilon_m\operatorname{sech}^2(a_m-r_h).
    \label{eq:schw-G}
\end{equation}
Without an environment, $G$ decreases monotonically with $r_h$ and has only one equilibrium zero, of charge $-1$; the environmental term can alter this monotonicity and introduce additional equilibrium branches.

At fixed $\tau=80$, solving $G=G_{,r_h}=G_{,r_hr_h}=0$ gives the amplitude threshold for interior folds:
\begin{equation}
    \epsilon_m^{\rm c}
    =
    \frac{3\sqrt{3}\pi}{2\tau}
    \simeq0.102026.
    \label{eq:schw-amplitude-threshold}
\end{equation}
This threshold can also be read from the radial derivative: the maximum positive environmental correction to $G_{,r_h}$ is $4\epsilon_m/(3\sqrt{3})$, and degenerate zeros become possible only when this reaches the magnitude $2\pi/\tau$ of the background slope.

\begin{figure}[htbp!]
    \centering
    \begin{subfigure}[t]{0.31\textwidth}
        \centering
        \includegraphics[width=\linewidth]{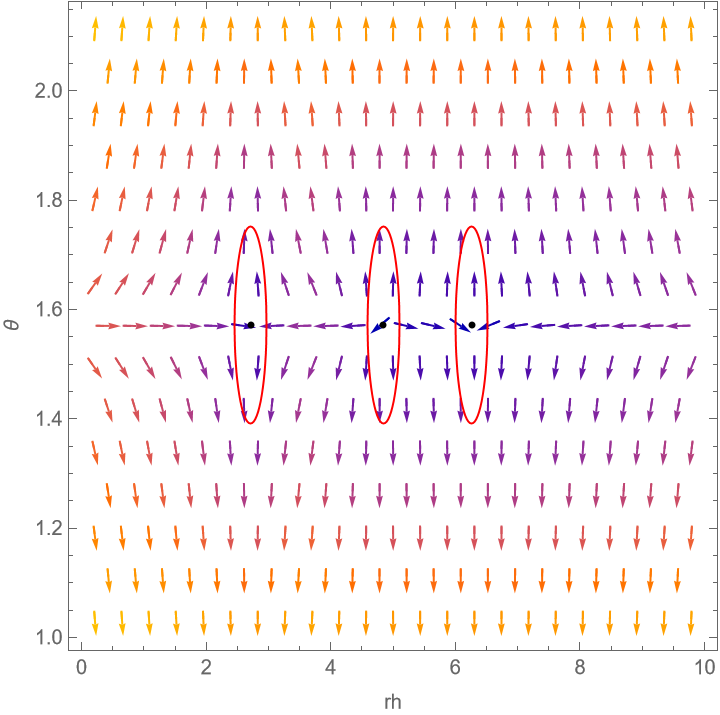}
        \caption{$a_m=3.5<a_m^{(1)}$}
        \label{fig:swz_am3.5}
    \end{subfigure}
    \hfill
    \begin{subfigure}[t]{0.31\textwidth}
        \centering
        \includegraphics[width=\linewidth]{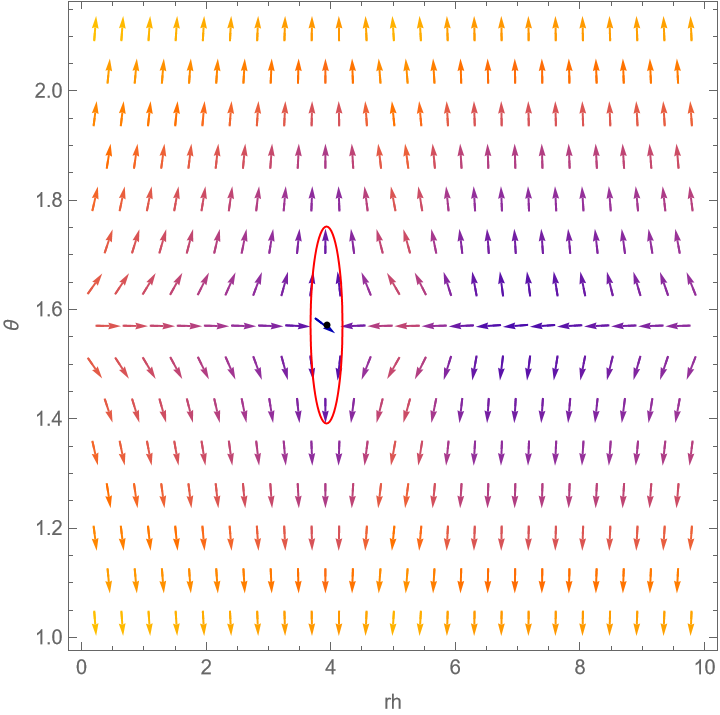}
        \caption{$a_m=5>a_m^{(1)}$}
        \label{fig:swz_am5}
    \end{subfigure}
    \hfill
    \begin{subfigure}[t]{0.31\textwidth}
        \centering
        \includegraphics[width=\linewidth]{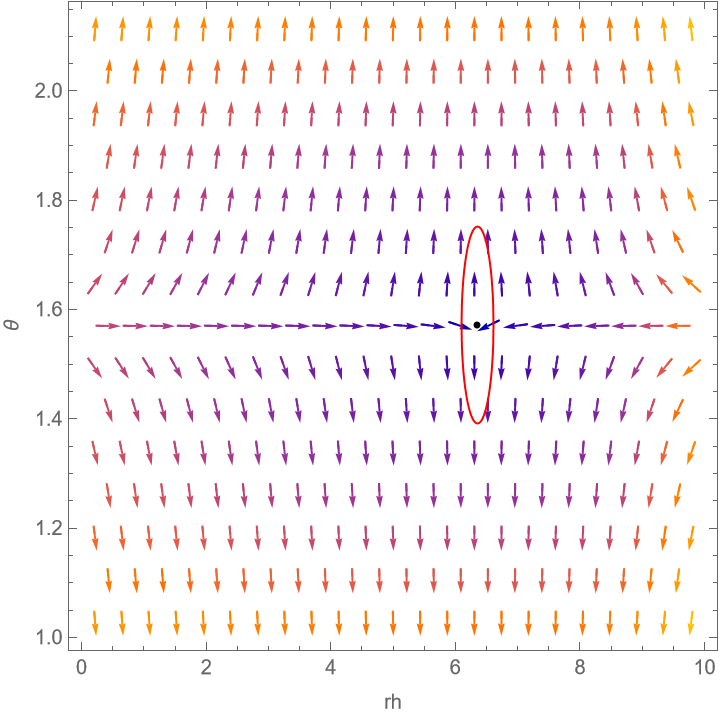}
        \caption{$a_m=10>a_m^{(1)}$}
        \label{fig:swz_am10}
    \end{subfigure}
    \captionsetup{width=0.9\textwidth}
    \caption{
    Topological vector fields of the Schwarzschild effective model at different environmental positions.
    The parameters are $\epsilon_m=0.5$ and $\tau=80$.
    }
    \label{fig:swz}
\end{figure}

We take
\begin{equation}
    \tau=80,
    \qquad
    \epsilon_m=0.5.
    \label{eq:schw-reference-parameters}
\end{equation}
The only interior fold point in $a_m>0$ is
\begin{equation}
    \left(r_h^{(1)},a_m^{(1)}\right)
    \simeq
    (5.844352,\,3.921665).
    \label{eq:schw-fold}
\end{equation}
Figure~\ref{fig:swz} shows the topological vector fields on either side of this critical position, with the corresponding zeros and charges listed in Table~\ref{tab:schw_charges}.

\begin{table}[htbp!]
    \centering
    \captionsetup{width=0.9\textwidth}
    \caption{
    Zero locations, topological charges, and total topological number for the Schwarzschild effective model.
    The parameters are $\epsilon_m=0.5$ and $\tau=80$.
    Zeros are ordered by increasing radius, with locations rounded to three decimal places.
    Charges are determined from Eq.~\eqref{eq:wi-analytic}.
    }
    \label{tab:schw_charges}
    \begin{tabular}{@{}lccc@{}}
        \toprule
        $a_m$ & Zeros $r_i$ & Charges $(w_i)$ & $W$\\
        \midrule
        $3.5$ & $2.719,\ 4.842,\ 6.266$
              & $-1,\ +1,\ -1$ & $-1$\\
        $5$   & $3.938$ & $-1$ & $-1$\\
        $10$  & $6.349$ & $-1$ & $-1$\\
        \bottomrule
    \end{tabular}
\end{table}

For $0<a_m<a_m^{(1)}$, the system has three simple zeros with ordered charges $(-1,+1,-1)$.
As $a_m$ increases, the intermediate stable branch and the large-radius unstable branch merge and annihilate at the point in Eq.~\eqref{eq:schw-fold}, while the smaller-radius unstable branch survives; interior reconstruction on this parameter slice therefore changes the zero count as $3\rightarrow1$.

The total topological number can be checked from the boundary behavior.
For the chosen parameters,
\begin{equation}
    G(0^+)
    =
    \frac{1}{2}\tanh^2a_m>0,
    \qquad
    G(\infty)<0.
    \label{eq:schw-boundary-signs}
\end{equation}
Both boundary signs remain unchanged for $a_m>0$, so the interior annihilation preserves $W=-1$, corresponding to class $W^{1-}$ in Sec.~\ref{sec:topological-classification}.

\begin{figure}[htbp!]
    \centering
    \includegraphics[width=0.62\linewidth]{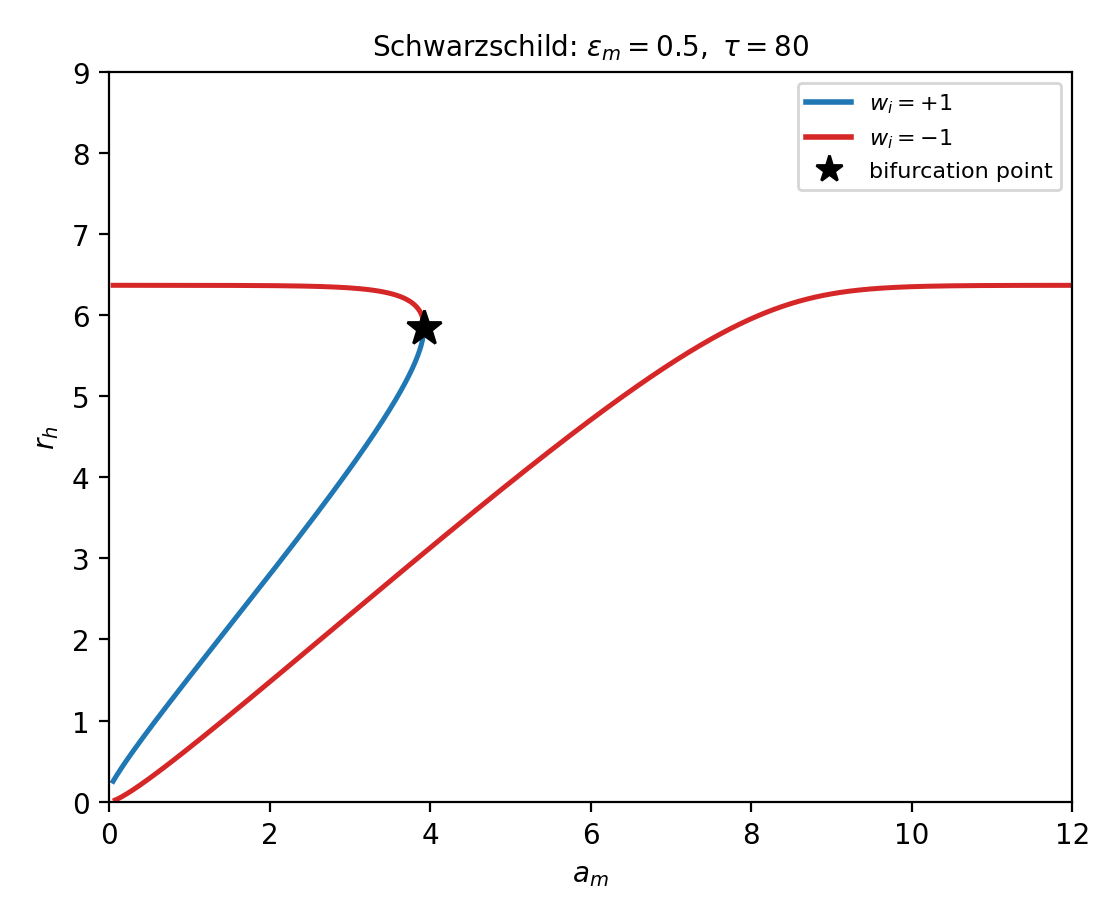}
    \captionsetup{width=0.9\textwidth}
    \caption{
    Equilibrium branches of the Schwarzschild effective model in the $(a_m,r_h)$ plane, with $\epsilon_m=0.5$ and $\tau=80$.
    The black star marks the interior fold point in $a_m>0$.
    }
    \label{fig:schw_phase}
\end{figure}

Figure~\ref{fig:schw_phase} traces the connections between these branches.
The environmental energy ratio at the critical point is
\begin{equation}
    \eta\left(r_h^{(1)},a_m^{(1)}\right)
    \simeq7.16\times10^{-3}.
    \label{eq:schw-critical-eta}
\end{equation}
Thus the localized energy term can change the extrema of the effective free energy even when the total environmental energy is much smaller than the background BH mass: a small energy ratio does not mean the local correction to the free-energy derivatives is negligible.

\subsection{Topological structure of the Reissner-Nordstr\"om black hole}

At fixed nonzero charge $Q$, the mass and Hawking temperature of the RN BH are
\begin{equation}
    M(r_h)
    =
    \frac{r_h^2+Q^2}{2r_h},
    \qquad
    T_H(r_h)
    =
    \frac{r_h^2-Q^2}{4\pi r_h^3}.
    \label{eq:rn-background-data}
\end{equation}
Nonextremal outer horizons satisfy $r_h>|Q|$.
The corresponding radial component of the topological vector field is
\begin{equation}
    G
    =
    \frac{1}{2}
    -\frac{Q^2}{2r_h^2}
    -\frac{2\pi r_h}{\tau}
    -\epsilon_m\operatorname{sech}^2(a_m-r_h).
    \label{eq:rn-G}
\end{equation}
The charge term changes the equilibrium structure at small radii, allowing the background to support a small-radius stable branch.

We choose
\begin{equation}
    \tau=80,
    \qquad
    \epsilon_m=0.5,
    \qquad
    Q=1.
    \label{eq:rn-reference-parameters}
\end{equation}
The interior degeneracy conditions give two critical points:
\begin{align}
    \left(r_h^{(1)},a_m^{(1)}\right)
    &\simeq
    (1.450709,\,2.666717),
    \label{eq:rn-first-fold}\\
    \left(r_h^{(2)},a_m^{(2)}\right)
    &\simeq
    (5.685101,\,3.723324).
    \label{eq:rn-second-fold}
\end{align}
Both lie within the physically admissible domain of nonextremal outer horizons.
At fixed $Q=1$ and $\tau=80$, the cusp condition $G=G_{,r_h}=G_{,r_hr_h}=0$ at low amplitude gives the minimum threshold for interior folds,
\begin{equation}
    \epsilon_m^{\rm c}\simeq0.093513,
    \label{eq:rn-amplitude-threshold}
\end{equation}
and the analysis below uses the amplitude slice specified in Eq.~\eqref{eq:rn-reference-parameters}.

\begin{figure}[htbp!]
    \centering
    \begin{subfigure}[t]{0.31\textwidth}
        \centering
        \includegraphics[width=\linewidth]{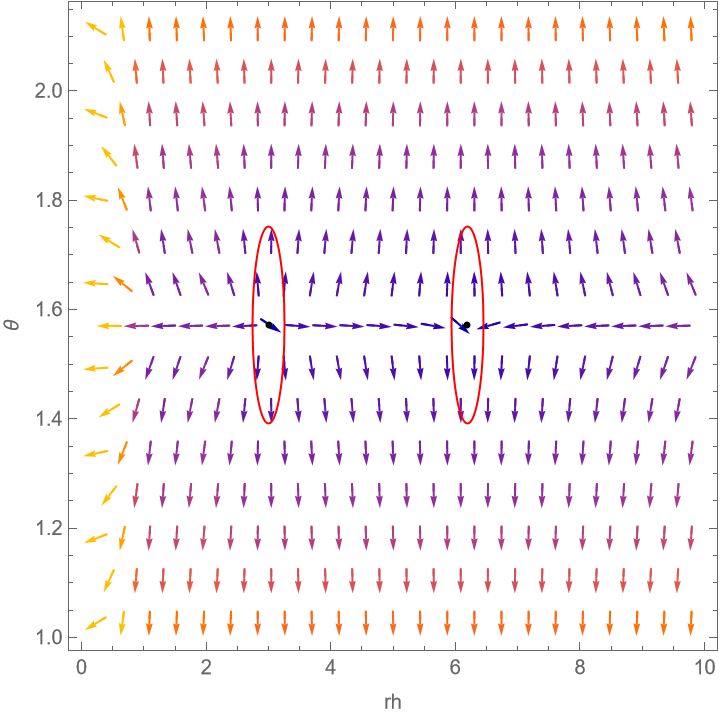}
        \caption{$a_m=2<a_m^{(1)}$}
        \label{fig:rn_am2}
    \end{subfigure}
    \hfill
    \begin{subfigure}[t]{0.31\textwidth}
        \centering
        \includegraphics[width=\linewidth]{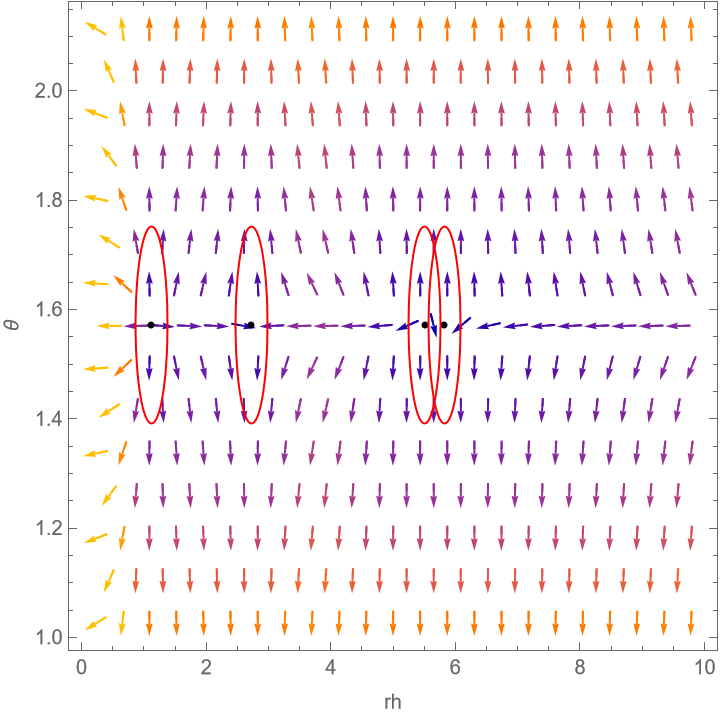}
        \caption{$a_m^{(1)}<3.7<a_m^{(2)}$}
        \label{fig:rn_am37}
    \end{subfigure}
    \hfill
    \begin{subfigure}[t]{0.31\textwidth}
        \centering
        \includegraphics[width=\linewidth]{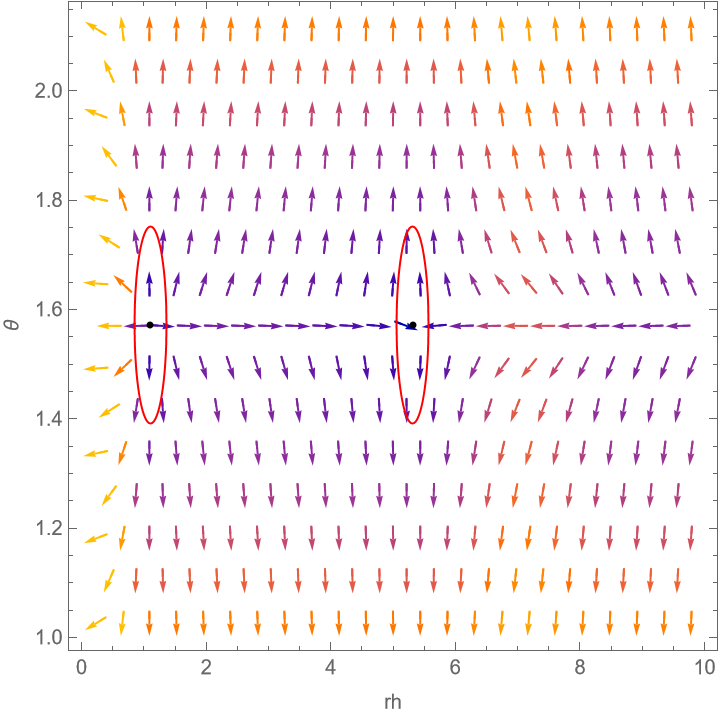}
        \caption{$a_m=7>a_m^{(2)}$}
        \label{fig:rn_am7}
    \end{subfigure}
    \captionsetup{width=0.9\textwidth}
    \caption{
    Topological vector fields of the RN effective model at different environmental positions.
    The parameters are $\epsilon_m=0.5$, $Q=1$, and $\tau=80$.
    }
    \label{fig:rn}
\end{figure}

Figure~\ref{fig:rn} and Table~\ref{tab:rn_charges} show the vector fields and zero structures for three representative configurations, respectively.

\begin{table}[htbp!]
    \centering
    \captionsetup{width=0.9\textwidth}
    \caption{
    Zero locations, topological charges, and total topological number for the RN effective model.
    The parameters are $\epsilon_m=0.5$, $Q=1$, and $\tau=80$.
    Zeros are ordered by increasing radius, with locations rounded to three decimal places.
    Charges are determined from Eq.~\eqref{eq:wi-analytic}.
    }
    \label{tab:rn_charges}
    \begin{tabular}{@{}lccc@{}}
        \toprule
        $a_m$ & Zeros $r_i$ & Charges $(w_i)$ & $W$\\
        \midrule
        $2$
        & $3.004,\ 6.195$
        & $+1,\ -1$ & $0$\\
        $3.7$
        & $1.117,\ 2.727,\ 5.511,\ 5.827$
        & $+1,\ -1,\ +1,\ -1$ & $0$\\
        $7$
        & $1.099,\ 5.318$
        & $+1,\ -1$ & $0$\\
        \bottomrule
    \end{tabular}
\end{table}

At the first critical position, a pair with charges $(+1,-1)$ appears at smaller radii than the two original branches, increasing the number of zeros from two to four.
As $a_m$ increases further, the original stable branch and the large-radius unstable branch merge and annihilate at the second critical point, while the two branches created in the first event survive — the process replaces equilibrium branches rather than annihilating the newly created pair with each other.

At the boundaries of the physical domain, the RN background satisfies
\begin{equation}
    G(|Q|)
    =
    -\frac{2\pi|Q|}{\tau}
    -\epsilon_m\operatorname{sech}^2(a_m-|Q|)
    <0,
    \qquad
    G(\infty)<0.
    \label{eq:rn-boundary-signs}
\end{equation}
The signs at both boundaries are the same, giving $W=0$ for each nondegenerate configuration; when equilibria exist, the smallest-radius branch is stable and the largest-radius branch is unstable, corresponding to class $W^{0+}$.

\begin{figure}[htbp!]
    \centering
    \includegraphics[width=0.62\linewidth]{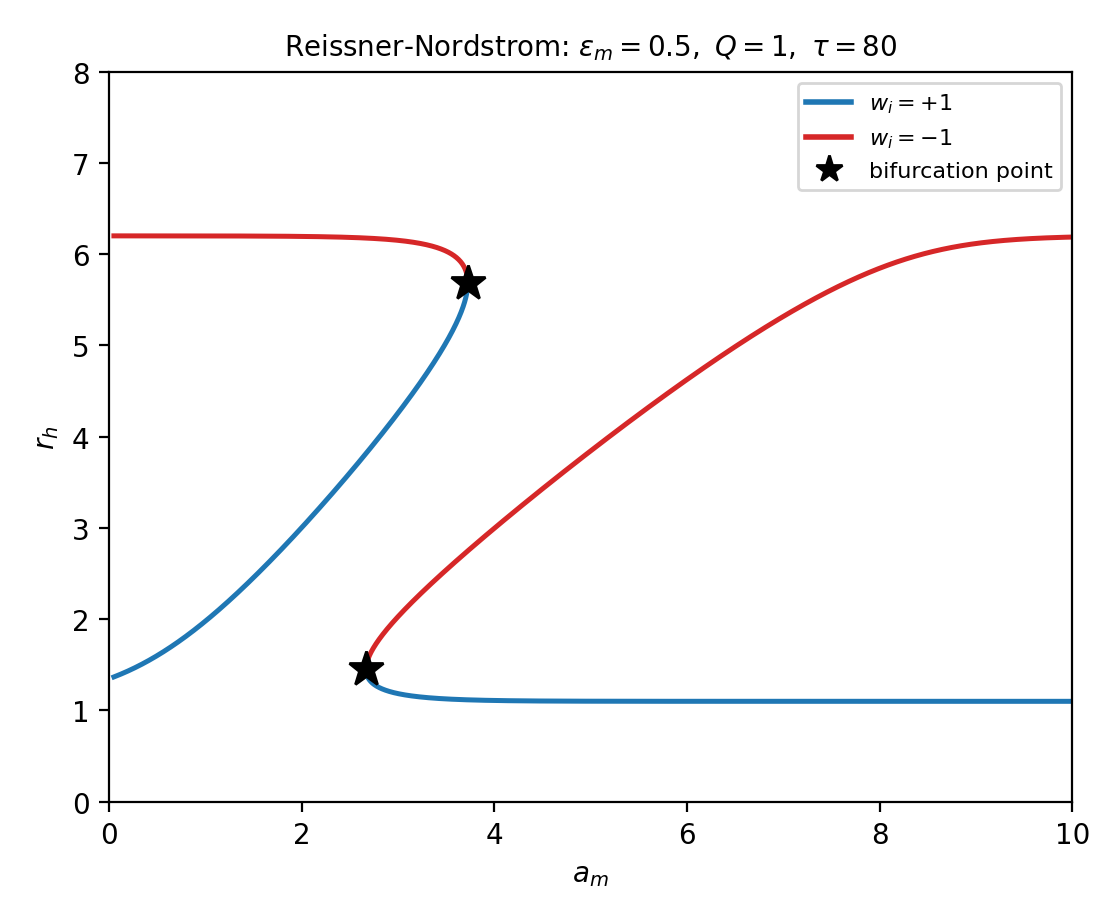}
    \captionsetup{width=0.9\textwidth}
    \caption{
    Equilibrium branches of the RN effective model in the $(a_m,r_h)$ plane, with $\epsilon_m=0.5$, $Q=1$, and $\tau=80$.
    Black stars mark the two interior fold points.
    }
    \label{fig:rn_phase}
\end{figure}

Figure~\ref{fig:rn_phase} shows the two sets of branches and their folds.
The energy ratios at the two critical points are
\begin{equation}
    \eta\left(r_h^{(1)},a_m^{(1)}\right)
    \simeq0.8591,
    \qquad
    \eta\left(r_h^{(2)},a_m^{(2)}\right)
    \simeq6.62\times10^{-3}.
    \label{eq:rn-critical-eta}
\end{equation}
At the same amplitude $\epsilon_m$, the relative environmental energies at the two reconstruction events differ by more than two orders of magnitude — a difference that follows from how the integrated exterior energy and the background mass jointly depend on the critical positions.
Thus $\epsilon_m$ and $\eta$ measure different properties of the environment.

\subsection{Topological structure of the Reissner-Nordstr\"om-AdS black hole}

The mass and Hawking temperature of the RN-AdS BH are
\begin{equation}
    M(r_h)
    =
    \frac{r_h}{2}
    +\frac{Q^2}{2r_h}
    +\frac{r_h^3}{2l^2},
    \qquad
    T_H(r_h)
    =
    \frac{1-Q^2/r_h^2+3r_h^2/l^2}{4\pi r_h}.
    \label{eq:rnads-background-data}
\end{equation}
Nonextremal outer horizons satisfy $r_h>r_{\rm e}$, where the extremal radius is given by
\begin{equation}
    r_{\rm e}^2
    =
    \frac{l^2}{6}
    \left(
        \sqrt{1+\frac{12Q^2}{l^2}}-1
    \right).
    \label{eq:rnads-extremal-radius}
\end{equation}
The corresponding radial component of the topological vector field is
\begin{equation}
    G
    =
    \frac{1}{2}
    -\frac{Q^2}{2r_h^2}
    +\frac{3r_h^2}{2l^2}
    -\frac{2\pi r_h}{\tau}
    -\epsilon_m\operatorname{sech}^2(a_m-r_h).
    \label{eq:rnads-G}
\end{equation}
Compared with the RN background, the AdS curvature term makes $G$ positive at large radii and changes the stability of the outermost branch.

\begin{figure}[htbp!]
    \centering
    \begin{subfigure}[t]{0.31\textwidth}
        \centering
        \includegraphics[width=\linewidth]{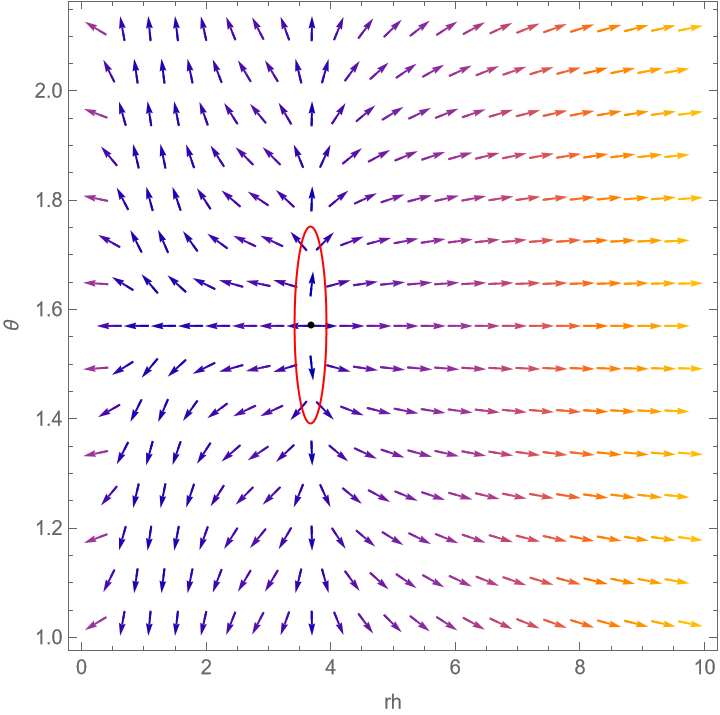}
        \caption{$a_m=3<a_m^{(1)}$}
        \label{fig:rnads_am3}
    \end{subfigure}
    \hfill
    \begin{subfigure}[t]{0.31\textwidth}
        \centering
        \includegraphics[width=\linewidth]{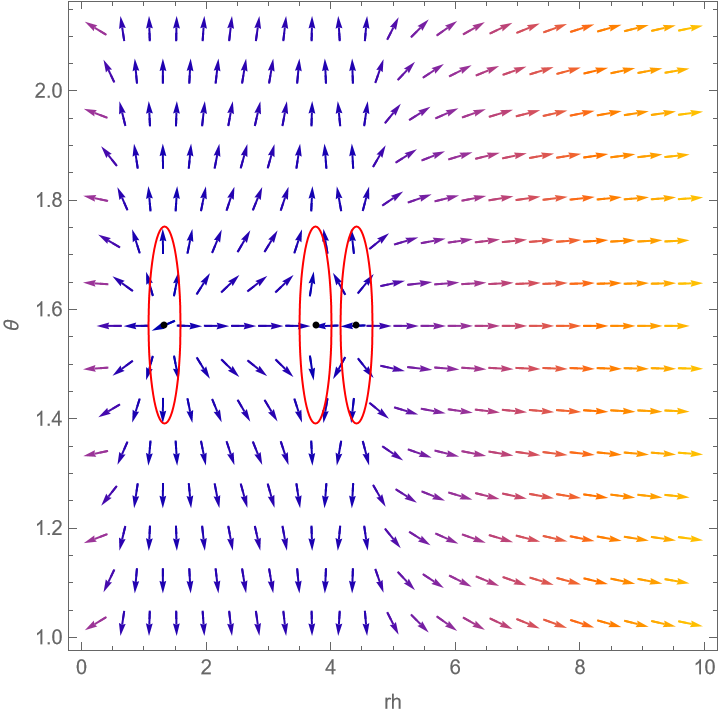}
        \caption{$a_m^{(1)}<4.4<a_m^{(2)}$}
        \label{fig:rnads_am44}
    \end{subfigure}
    \hfill
    \begin{subfigure}[t]{0.31\textwidth}
        \centering
        \includegraphics[width=\linewidth]{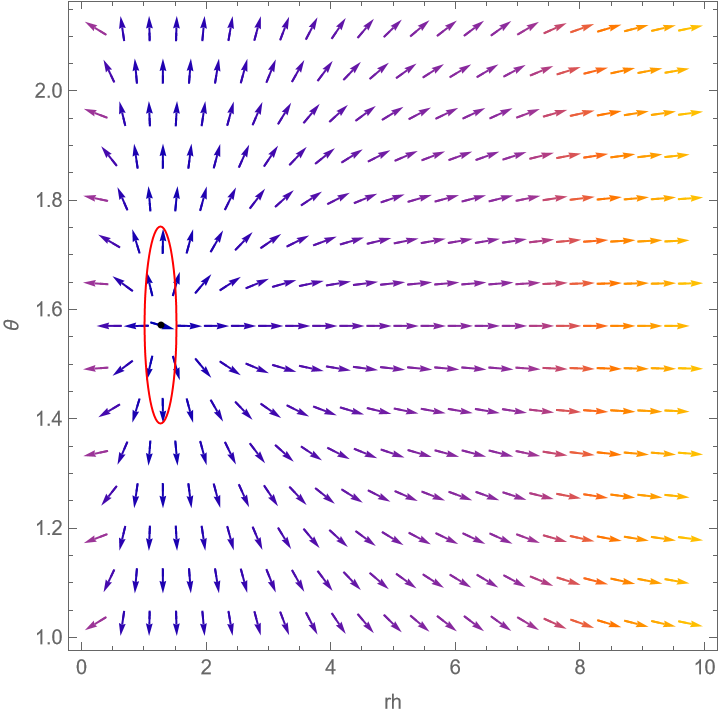}
        \caption{$a_m=7>a_m^{(2)}$}
        \label{fig:rnads_am7}
    \end{subfigure}
    \captionsetup{width=0.9\textwidth}
    \caption{
    Topological vector fields of the RN-AdS effective model at different environmental positions.
    The parameters are $\epsilon_m=5$, $Q=1$, $l=2$, and $\tau=10$.
    }
    \label{fig:rnads}
\end{figure}

We choose
\begin{equation}
    \tau=10,
    \qquad
    \epsilon_m=5,
    \qquad
    Q=1,
    \qquad
    l=2.
    \label{eq:rnads-reference-parameters}
\end{equation}
The extremal radius is then $r_{\rm e}=\sqrt{2/3}\simeq0.816497$.
Numerically solving the interior degeneracy conditions yields
\begin{align}
    \left(r_h^{(1)},a_m^{(1)}\right)
    &\simeq
    (1.861600,\,3.695405),
    \label{eq:rnads-first-fold}\\
    \left(r_h^{(2)},a_m^{(2)}\right)
    &\simeq
    (4.262509,\,4.550332).
    \label{eq:rnads-second-fold}
\end{align}
Both critical points satisfy $r_h>r_{\rm e}$.
Figure~\ref{fig:rnads} and Table~\ref{tab:rnads_charges} show representative zero structures on either side of the critical positions.

\begin{table}[htbp!]
    \centering
    \captionsetup{width=0.9\textwidth}
    \caption{
    Zero locations, topological charges, and total topological number for the RN-AdS effective model.
    The parameters are $\epsilon_m=5$, $Q=1$, $l=2$, and $\tau=10$.
    Zeros are ordered by increasing radius, with locations rounded to three decimal places.
    Charges are determined from Eq.~\eqref{eq:wi-analytic}.
    }
    \label{tab:rnads_charges}
    \begin{tabular}{@{}lccc@{}}
        \toprule
        $a_m$ & Zeros $r_i$ & Charges $(w_i)$ & $W$\\
        \midrule
        $3$   & $3.684$
              & $+1$ & $1$\\
        $4.4$ & $1.327,\ 3.760,\ 4.411$
              & $+1,\ -1,\ +1$ & $1$\\
        $7$   & $1.274$
              & $+1$ & $1$\\
        \bottomrule
    \end{tabular}
\end{table}

\begin{figure}[htbp!]
    \centering
    \includegraphics[width=0.62\linewidth]{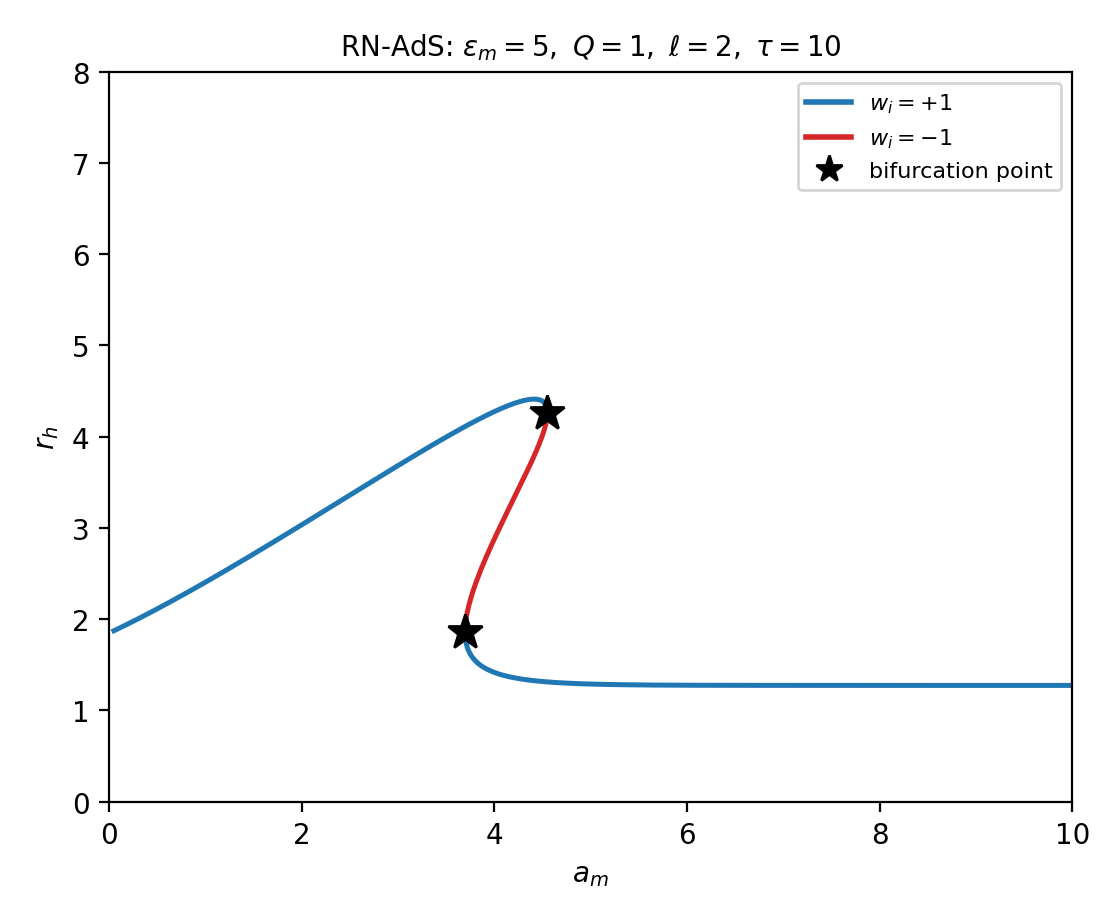}
    \captionsetup{width=0.9\textwidth}
    \caption{
    Equilibrium branches of the RN-AdS effective model in the $(a_m,r_h)$ plane, with $\epsilon_m=5$, $Q=1$, $l=2$, and $\tau=10$.
    Black stars mark the two interior fold points.
    }
    \label{fig:rnads_phase}
\end{figure}

For $a_m<a_m^{(1)}$, there is only one stable zero.
Beyond $a_m^{(1)}$, a stable-unstable pair appears at smaller radii, giving the charge sequence $(+1,-1,+1)$.
At the second critical point, the new unstable branch merges with the original large-radius stable branch and annihilates, leaving the new small-radius stable branch, so the number of zeros changes as $1\rightarrow3\rightarrow1$ — the branch connections are shown in Fig.~\ref{fig:rnads_phase}.

At the lower boundary of the physical domain, the extremality condition gives
\begin{equation}
    G(r_{\rm e})
    =
    -\frac{2\pi r_{\rm e}}{\tau}
    -\epsilon_m\operatorname{sech}^2(a_m-r_{\rm e})
    <0,
    \qquad
    G(\infty)>0,
    \label{eq:rnads-boundary-signs}
\end{equation}
so the effective model preserves $W=+1$ throughout this interior reconstruction, with both the smallest- and largest-radius branches stable — class $W^{1+}$.

To determine the onset of reconstruction as the amplitude varies, we fix $\tau=10$, $Q=1$, and $l=2$ and solve the cusp condition $G=G_{,r_h}=G_{,r_hr_h}=0$, which gives
\begin{equation}
    \left(
        r_h^{\rm c},
        a_m^{\rm c},
        \epsilon_m^{\rm c}
    \right)
    \simeq
    (2.190002,\,3.015195,\,1.513999).
    \label{eq:rnads-amplitude-threshold}
\end{equation}
The environmental energy ratio at this point is
\begin{equation}
    \eta_{\rm c}\simeq0.9636.
    \label{eq:rnads-cusp-eta}
\end{equation}
For the illustrative amplitude $\epsilon_m=5$, the two fold points instead give
\begin{equation}
    \eta\left(r_h^{(1)},a_m^{(1)}\right)
    \simeq4.8613,
    \qquad
    \eta\left(r_h^{(2)},a_m^{(2)}\right)
    \simeq0.5366.
    \label{eq:rnads-critical-eta}
\end{equation}
These values show that, near the threshold and the critical configurations displayed here, the environmental energy is not much smaller than the BH mass: the result characterizes reconstruction in the strong-environment regime of the effective free energy, whose physical realization requires incorporating environmental backreaction on the geometry.
This assessment applies to the chosen parameters; it does not imply that local reconstruction is harder for RN-AdS BHs than for other backgrounds under all parameter choices.

A question for further study is whether thermodynamic branch reconstruction in a self-consistent matter environment is related to shell mechanical stability or critical changes in the horizon structure.
Nonvacuum BH models provide a basis for treating the environment and geometry consistently~\cite{Cardoso:2021wlq,Visser:1992qh}, and thin matter shells and anisotropic fluid models have also been used to study environmental effects on BH responses~\cite{Cardoso:2019upw}.
If the smooth environmental profile is replaced by a self-gravitating thin shell, the Israel junction conditions can relate the shell stresses to the interior and exterior geometries~\cite{Israel:1966rt}.
Combining this construction with charged thin-shell thermodynamics~\cite{Lemos:2015ama} and the canonical ensemble for AdS thin shells~\cite{Fernandes:2025shell} would allow comparisons of mechanical equilibrium, thermodynamic stability, and topological bifurcations.
We have not established a correspondence between these critical behaviors in the present study.

\subsection{Comparison of black holes in different topological classes}

The four backgrounds share the localized environmental term in Eq.~\eqref{eq:master-G}; their differences come from the radial derivative of the background free energy and the boundaries of the physical domain.
Tables~\ref{tab:sads_charges}, \ref{tab:schw_charges}, \ref{tab:rn_charges}, and \ref{tab:rnads_charges} show that opposite-charge pairs can be created or annihilated in different initial topological classes within the chosen interior reconstruction intervals, and Table~\ref{tab:background-comparison} summarizes the branch changes and critical energy ratios.

\begin{table}[htbp!]
    \centering
    \captionsetup{width=0.9\textwidth}
    \caption{
    Interior topological reconstruction in the four backgrounds for the illustrative parameters used here.
    Zero-count changes are listed in order of increasing $a_m$, excluding the degenerate points themselves.
    Energy ratios are ordered by the corresponding critical values of $a_m$.
    The SAdS row is restricted to $a_m>a_{\rm b}$, and the Schwarzschild row to $a_m>0$.
    }
    \label{tab:background-comparison}
    \begin{tabular}{@{}lcccc@{}}
        \toprule
        Background
        & Class
        & $W$
        & Zero-count change
        & Critical energy ratios $\eta$\\
        \midrule
        SAdS
        & $W^{0-}$ & $0$
        & $2\rightarrow4\rightarrow2$
        & $0.2420,\ 0.1686$\\
        Schwarzschild
        & $W^{1-}$ & $-1$
        & $3\rightarrow1$
        & $0.00716$\\
        RN
        & $W^{0+}$ & $0$
        & $2\rightarrow4\rightarrow2$
        & $0.8591,\ 0.00662$\\
        RN-AdS
        & $W^{1+}$ & $+1$
        & $1\rightarrow3\rightarrow1$
        & $4.8613,\ 0.5366$\\
        \bottomrule
    \end{tabular}
\end{table}

These examples show that branch reconstruction induced by a localized environment is not restricted to a particular initial topological class: the same total topological number can correspond to different zero counts and stability orderings, so global classification and local defect structure must be examined separately~\cite{Chen:2024MultipleDefects}.
Here, topological conservation refers to interior reconstruction with unchanged boundary behavior — the boundary crossings at smaller $a_m$ in the Schwarzschild-AdS background were discussed separately above.

The critical positions and required environmental contributions depend strongly on the background and parameters.
The illustrative Schwarzschild fold occurs at a small environmental energy ratio, the two RN events span different energy regimes, and the displayed RN-AdS critical configurations require a substantial environmental energy contribution.
Because the models use different $\tau$ and $\epsilon_m$, these results illustrate possible differences in response without establishing a universal ranking of robustness across backgrounds; a systematic comparison would need a common parameter-scan scheme that examines amplitude thresholds, critical positions, window widths, and energy ratios together.

\section{Conclusions}
\label{sec:conclusions}

We have shown that, in an effective fixed-background description, the position of a localized environment can change the number and stability ordering of BH equilibrium branches without necessarily changing the total topological number.
Taking the peak position $a_m$ as the control parameter, we found this local topological reconstruction in Schwarzschild-AdS, Schwarzschild, Reissner-Nordstr\"om, and Reissner-Nordstr\"om-AdS BHs.
The total topological number alone therefore cannot characterize the thermodynamic response to an environment.
The number and positions of the local defects, and the way they connect into branches, must be tracked as well.

The mechanism is local.
The environment adds a localized correction to the slope and curvature of the free energy, and varying $a_m$ moves this correction along the radial axis.
Where the degenerate-zero conditions hold, two neighboring zeros of opposite charge merge or split in a fold bifurcation.
In the Schwarzschild-AdS BH, two critical positions bound a window of additional branches, in which the number of zeros grows from two to four and returns to two.
The pair annihilated at the second critical position is not the pair created at the first; the new stable branch replaces the original large-radius one.
The total topological number is conserved as long as no zero crosses the boundary of the physical domain.
When one does, the number changes: in the Schwarzschild-AdS example, from $+1$ to $0$ as $a_m$ increases through $a_{\rm b}$, where the small-radius zero reaches $r_h=0$.

The critical positions, amplitude thresholds, and required environmental energies differ among the backgrounds.
At the Schwarzschild fold, the ratio of environmental energy to BH mass is $\eta\simeq7.16\times10^{-3}$, so an environment carrying less than $1\%$ of the mass can still shift the free-energy extrema.
In the RN-AdS example, $\eta\simeq0.964$ already at the amplitude threshold, so reconstruction requires environmental energy comparable to the BH mass.
The two RN critical events differ in $\eta$ by more than two orders of magnitude at the same $\epsilon_m$, so amplitude and relative energy are not interchangeable measures.
Because the examples use different $\tau$ and $\epsilon_m$, they do not rank the backgrounds by robustness.

Our results rely on an effective free energy that adds the environmental energy to the BH free energy.
They are not self-consistent thermal equilibria of a BH and its matter environment: we neglect backreaction, the environmental entropy, the work needed to hold the environment in place, and the Tolman redshift of the local temperature.
The strong-environment cases, RN-AdS in particular, therefore need to be tested in a coupled model.

Two extensions follow.
The first is a coupled model, with a self-gravitating shell or a continuous matter distribution, in which the geometry, the temperature redshift, and the thermodynamic potential are determined together.
It would show how much of the reconstruction survives, and whether the thermodynamic bifurcations are linked to the mechanical stability of the matter or to changes in the horizon structure; the energy ratio alone cannot establish such a link.
The second is dynamical.
The branch diagrams cannot tell which state the system reaches when a stable extremum disappears, or how long the transition takes.
Once the control process and relaxation mechanism are specified, the nonequilibrium evolution near the critical points can be studied, connecting the static reconstruction found here to the state changes a system actually undergoes. 

\section*{Acknowledgement}
Z.-M. Xu is supported by the National Natural Science Foundation of China (Grant Nos. 12575064, 12247103), by the Natural Science Basic Research Plan in Shaanxi Province of China (Grant No. 2025JC-YBQN-029), and by the project of Tang Scholar in Northwest University.

\appendix

\bibliographystyle{utphys}
\bibliography{references}

\end{document}